*Article*

# Photokinetics of photothermal reactions

Mounir MAAFI

Leicester School of Pharmacy, De Montfort University, The Gateway, Leicester, LE1 9BH and UK, *mmaafi@dmu.ac.uk*

**Abstract:** Photothermal reactions, involving both photochemical and thermal reaction-steps, are the most abundant sequences in photochemistry. The derivation of their rate-laws is standardized, but the integration of these rate-laws has not yet been achieved. Indeed, the field still lacks integrated rate-laws for the description of these reactions' behavior, and/or identification of their reaction-order. This made a comprehensive account of the photokinetics of photothermal reactions to be a gap in the knowledge. This gap is addressed in the present paper by introducing an unprecedented general model equation capable to mapping out the kinetic traces of such reactions when exposed to light or in the dark. The integrated rate-law model equation also applies when the reactive medium is exposed to either monochromatic or polychromatic light irradiation. The validity of the model equation was established against simulated data obtained by a fourth-order Runge-Kutta method. It was then used to describe and quantify several situations of photothermal reactions, such as the effects of initial concentration, spectator molecules, and incident radiation intensity, and the impact of the latter on the photonic yield. The model equation facilitated a general elucidation method to determine the intrinsic reaction parameters (quantum yields and absorptivities of the reactive species) for any photothermal mechanism whose number of species are known. This paper contributes to rationalising photokinetics along the same general guidelines adopted in chemical kinetics.

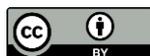

## 1. Introduction

The generic term of "photoreaction" is equally employed, in the literature, to designate reactions involving exclusively pure photochemical reaction-steps, and those which undergo both photochemical and thermal processes. A subtilty that might suggest to assign the terminology "photothermal reactions" for the latter group of reactions, leaving the term photoreactions to be used for the former type of sequences. Kinetically, such a distinction makes sense because thermal and photo-processes are differently quantified. For simplicity, the reactivity concerned here is, overall, supposed to evolve within a timeframe compatible with observation by spectrophotometry, a technique commonly used in and required for the kinetic studies of such reactions (even though, light excitation and most of the subsequent events it induces, are much faster processes to be detected by routine spectroscopic techniques [1]).

The most ubiquitous type of light induced reactions consists of photothermal reactions. Photothermal materials have shown potential for application in many technologically important fields [2-8]. In recent years, this family of chemical compounds has known remarkable developments and a diversity of applications, including in mechanochemical reactivity [9], mechanochromism [10], chemical sensing [11,12], photochromes grafted DNA [13], biological imaging [14,15], drug delivery systems [16], microreactors [17], polymers [18], rewritable data-storage [19], photochromes [20,21], therapeutics [22], environmental issues [23,24], optical devices [25], nano-materials[26,27], photocatalysis [28], photoswitchable nanoclusters [29], actinometry [30], green photochemistry [31], scaling up of photoreaction [32], and more [8,33,34].

Current and future development of photothermal materials is supposed to not only be achieved by the design of new chemical structures but also by a better assessment, quantification and control of the reaction properties. The former aspect is mainly realized by preparative organic chemistry, whereas the latter is achievable by photokinetics (kinetics of photo- and photothermal reactions). Though, its striking to observe that the investigative photokinetic tools have historically known an overall relatively poor development. For example, conversely to thermal reactions which are characterised by defined kinetic orders, such as $0^{th}$-, $1^{st}$-, and $2^{nd}$-order kinetics, the order of a photoreaction is still not known within the wider photochemistry literature [2,3,5-8,35-37]. This fact also meant that the integrated rate-laws, expressing species concentrations as functions of reaction time ($C = f(t)$), which are essential for investigation and quantification, are likewise unknown. In general terms, the lack of photokinetic tools for photo- and photothermal reactions originally stems from a difficulty or impossibility to analytically solve the systems of differential equations that fundamentally describe photothermal reaction kinetics.

Consequently, it is ubiquitous in the literature that kinetic treatments suitable for thermal relaxation (e.g., depicted by mono- or bi-exponential functions) are applied for analyses of data collected on photothermal reactions. For example, no specific photokinetic treatment was proposed for the data corresponding to photochromes kinetic data [20,31]. In the same context, the quantum yield of the reactant has usually been estimated by dividing the remaining concentration of that species after a given reaction time ($C(0) - C(t)$) by the total number of photons of the radiation reaching the reactive medium in the same period (supposedly fully absorbed by the reactant). Such a procedure is most often employed irrespective on whether the reaction mechanism is fully known, and/or the reaction is subjected to mono- or poly- chromatic light [38-41]. Unfortunately, this fundamental parameter must be determined experimentally. An attempt to computing the quantum yield by theoretical analysis of the potential energy surface was proposed, but this area of research is still developing [42]. These features represent a few examples underlining the considerable limitation of the toolkit offered to an experimentalist investigating photo- and/or photothermal reaction kinetics.

This gap in the knowledge has, in part, been addressed in recent publications that attempted to palliate the lack of integrated rate-laws by introducing a general formula suitable to fit the kinetic traces of any photoreaction be it exposed to monochromatic or polychromatic light [43,44].

The present paper attempts to contribute to this effort by presenting an explicit model equation, that stands for an integrated rate-law for photothermal reactions, and exploit this tool to advance understanding, rationalization, and prediction of photokinetics of photothermal reactions.

## 2. Results and discussion
### 2.1. The rate-law

The kinetic behavior of a photothermal reaction involving $n_{sp} = v + 1$ species, evolving from the reactant ($X = Y_0$), is fully captured by the general integro-differential equation (Eq.1), that stands here for the rate-law ($r_{Y_j}^{Lp,\Delta\lambda,T}(t)$, in $mol\ dm^{-3}\ s^{-1}$) of any species $Y_j$ involved in the mechanism governing the $XY_v(q\Phi, uk)$ reaction.

Eq.1 describes a reaction medium maintained at a constant temperature ($T$) and subjected to a collimated polychromatic beam delivered by lamp $Lp$ and covering the wavelength domain $\Delta\lambda = \lambda_b - \lambda_a$ (expressed in $nm$), that overlaps with the reactive medium absorption spectrum (also dubbed the Overlap Section between Irradiation and Absorption, *OSIA* [44]).

The formulation of Eq.1 assumes that the concentration of the excited-state species, reflexion, emission and scattering are negligible, and the homogenous reactive medium is continuously and vigorously stirred.

Because of the dual photo- and thermal nature of the reaction, it is conjectured that its kinetics will be mapped out by a mix of both $\Phi$-order [45] and exponential type functions, translating, respectively, the kinetic behaviors of the photochemical and the thermal processes (*vide infra* Section 2.2).

$$r_{Y_j}^{Lp,\Delta\lambda,T}(t) = \frac{dC_{Y_j}^{Lp,\Delta\lambda,T}(t)}{dt} = r^{Lp,\Delta\lambda}(t) + r^{\Delta}(t)$$

$$= \int_{\lambda_a}^{\lambda_b} \left( \sum_{j';\, j'\neq j}^{n_{\Phi,j}} \left( -\Phi_{Y_j \to Y_{j'}}^{\lambda_{irr}} P_{a_{Y_j}}^{\lambda_{irr}}(t) + \Phi_{Y_{j'} \to Y_j}^{\lambda_{irr}} P_{a_{Y_{j'}}}^{\lambda_{irr}}(t) \right) \right) d\lambda \quad (1)$$

$$+ \sum_{j';\, j'\neq j}^{n_{\Delta,j}} \left( -k_{Y_j \to Y_{j'}}^{\Delta} C_{Y_j}^{Lp,\Delta\lambda,T}(t) + k_{Y_{j'} \to Y_j}^{\Delta} C_{Y_{j'}}^{Lp,\Delta\lambda,T}(t) \right)$$

The variation of species $Y_j$ concentration ($C_{Y_j}^{Lp,\Delta\lambda,T}(t)$, in $mol\ dm^{-3}$) with reaction time ($t$ in $s$) is due to a combination of purely photochemical and purely thermal contributions.

In Eq.1, the $r^{\Delta}(t)$ kinetics of the $n_{\Delta,j}$ thermal reaction-steps (starting or ending at $Y_j$: $Y_j \to Y_{j'}$ or $Y_{j'} \to Y_j$, with $Y_{j'}$ being an adjacent or a directly linked species to $Y_j$) is expressed by the usual first-order equations employed in thermal chemical kinetics, with a rate $r_{Y_j}^{\Delta}(t)$, depicted here as a product of the rate-constant of the considered reaction-step ($k_{Y_j \to Y_{j'}}^{\Delta}$ or $k_{Y_{j'} \to Y_{j'}}^{\Delta}$, in $s^{-1}$), and the concentration of the species at the start of that reaction-step ($C_{Y_j}^{Lp,\Delta\lambda,T}(t)$ or $C_{Y_{j'}}^{Lp,\Delta\lambda,T}(t)$, respectively).

The contribution of the $n_{\Phi,j}$ photochemical reaction-steps related to $Y_j$, $r^{Lp,\Delta\lambda}(t)$, is conveyed by an integral over $\Delta\lambda$ of a sum of products each involving the dimensionless quantum yield of the considered reaction-step (the reaction-step starting or ending at $Y_j$, $\Phi_{Y_j \to Y_{j'}}^{\lambda_{irr}}$ or $\Phi_{Y_{j'} \to Y_j}^{\lambda_{irr}}$), and the time-dependent amount of the light absorbed by the species at the start of the considered photochemical reaction-step ($P_{a_{Y_j\, or\, j'}}^{\lambda_{irr}}(t)$, in $eisntein\ dm^{-3}\ s^{-1}$). These quantities are defined at given individual irradiation wavelengths, $\lambda_{irr}$, of the *OSIA* $\Delta\lambda$-domain.

The light absorbed by a species at $\lambda_{irr}$, is worked out from the Beer-Lambert law (the derivation of this quantity was previously provided [43]), as

$$P_{a_{Y_j\, or\, j'}}^{\lambda_{irr}}(t) = \frac{A_{Y_j\, or\, j'}^{\lambda_{irr},T/\lambda_{irr}}(t)}{A_{tot}^{\lambda_{irr},T/\lambda_{irr}}(t)} P_0^{\lambda_{irr}} \left(1 - 10^{-A_{tot}^{\lambda_{irr},T/\lambda_{irr}}(t)}\right) = A_{Y_j\, or\, j'}^{\lambda_{irr},T/\lambda_{irr}}(t)\, P_0^{\lambda_{irr}}\, PKF^{\lambda_{irr}}(t) \quad (2)$$

where the dimensionless total absorbance at wavelength $\lambda_{irr}$ ($A_{tot}^{\lambda_{irr},T/\lambda_{irr}}(t)$, Eq.3), represents a sum of the individual absorbances of the $n_{sp}$ species present at time $t$ and absorbing at $\lambda_{irr}$ ($A_{Y_j}^{\lambda_{irr},T/\lambda_{irr}}(t)$). The incident number of photons, $P_0^{\lambda_{irr}}$, at the wavelength $\lambda_{irr}$, entering the reactor per second and per irradiated area and volume of the investigated sample, is expressed in $einstein\ s^{-1}\ dm^{-3}$.

$PKF(t)$ is the dimensionless and time-dependent photokinetic factor. In Eq.3, $\varepsilon_{Y_j}^{\lambda_{irr}}$ (in $mol^{-1}\ dm^3\ cm^{-1}$), is the absorptivity of species $Y_j$ at $\lambda_{irr}$, and $l_{irr}$ (in $cm$), the optical path length of the irradiation light from $Lp$ inside the reactive medium. Incidentally, the initial concentration of the reactant is always equal to the sum of the species concentrations present at any time $t$, for a given reaction (i.e., $C_X^{Lp,\Delta\lambda,T}(0) = \sum C_{Y_j}^{Lp,\Delta\lambda,T}(t)$). The total absorbance (Eq.3) is experimentally measured at an observation wavelength, $\lambda_{obs}$ (on the spectrum of the reactive medium), through the observation optical path length, $l_{obs}$, crossed by the monitoring light inside the sample. Both $\lambda_{obs}$ and $l_{obs}$ might or might not be equal to their irradiation counterparts $\lambda_{irr}$ and $l_{irr}$, respectively. For simplicity, in the following equations the optical path length are considered equal, i.e., the ratio $l_{irr}/l_{obs} = 1$.

$$A_{tot}^{\lambda_{irr},T/\lambda_{obs}}(t) = \sum_{j=0}^{n_{sp}} A_{Y_j}^{\lambda_{irr},T/\lambda_{obs}}(t) = \sum_{j=0}^{n_{sp}} \varepsilon_{Y_j}^{\lambda_{obs}} l_{irr} C_{Y_j}^{Lp,\Delta\lambda,T}(t) \tag{3}$$

The coherence of Eq.1 is relayed by the equality of the dimensions of, on one hand, the thermal ($mol\ dm^{-3}\ s^{-1}$) and phtochemical ($einstein\ s^{-1}\ dm^{-3}$) contributions in the right hand-side of the rate-equation (Eq.1, $r^{Lp,\Delta\lambda}(t)$ and $r^{\Delta}(t)$, respectively), since an *einstein* is a mole of photons, and on the other hand, to that of the differential on the left hand-side of the equation ($dC_{Y_j}^{Lp,\Delta\lambda,T}(t)/dt$, given in $mol\ dm^{-3}\ s^{-1}$).

Also, the validity of Eq.1 clearly requires a valid Eq.2, which in turn imposes that the absorbance data used for these formulae must obey the Beer-Lambert law, i.e., only data belonging to the linearity range of the absorbance calibration graph must be considered. This holds for the $n_{sp}$ species of the reactive medium, that absorb at any $\lambda_{irr}$ of the *OSIA* $\Delta\lambda$-domain ($A_{Y_j}^{\lambda_{irr}/\lambda_{obs},T}(t) \leq 0.5$).

For the purpose of generalization, the quantum yields (as well as all the parameters in Eq.2) are considered here to be wavelength-dependent. In addition, because the temporal variations of concentration and rate are intimately dependent on the *OSIA* wavelength-range, the profile of the lamp within *OSIA*, and the temperature of the medium, the concentration, $C_{Y_j}$, the total absorbance, $A_{tot}$, and the rate, $r_{Y_j}$, must be referenced to $Lp$, $\Delta\lambda$, and $T$.

The photochemical part of Eq.1 is non-zero only and only if all its parameters $\Phi$, $\varepsilon$, $P_0$, $l_{irr}$ and $C(0)$ have concomitantly non-zero values at a given wavelength $\lambda_{irr}$ (otherwise, no photochemical contribution is accounted for when any of these factors is equal to zero at $\lambda_{irr}$). We remark that Eq.1 reduces either to a classical differential equation of thermal reactions, $r^{\Delta}(t)$, if considered for a reactive system observed in the dark ($P_0 = 0$), or to the rate-law of a photochemical reaction, $r^{Lp,\Delta\lambda}(t)$ [44], in the absence of thermal reaction-steps ($k^{\Delta} = 0$).

Removing the integral over $\lambda$ from the photochemical part of Eq.1, generates the rate-law for a photothermal reaction when exposed to a monochromatic light beam (then the photochemical part in the modified Eq.1 will have the same formulation as previously stated [43]). Accordingly, Eq.1 can describe the three (thermal, photochemical, and photothermal) reactions considered either under mono- or polychromatic light irradiations.

If Eq.1 describes fully the physical system, this equation remains, however, impossible to be mathematically solved analytically. Also, and contrasting with many integro-differential equations whose analytical integrations are standardized [46,47], the kind of Eq.1 has not even benefited from any research proposing its evaluation by numerical methods. The difficulty here probably stems from the many different and complicate functions describing the wavelength-dependent coefficients of the $r^{Lp,\Delta\lambda}(t)$ term in Eq.1 (most often these functions are specific to each experiment and their explicit formulae are not known – except by proceeding to deconvolution of their experimentally obtained envelops).

As a way out of this unfortunate situation, it was proposed to introduce an assumption whereby the integral in Eq.1 is considered equivalent to a sum at a 1-nm step [44,48]. Accordingly, Eq.1 transforms into Eq.4, and hence can readily be evaluated by numerical integration methods as the fourth-order Runge-Kutta method employed in the present work.

$$Theo: r_{Y_j}^{Lp,\Delta\lambda,T}(t) = \sum_{\lambda_{irr}=\lambda_a}^{\lambda_b} \left( \sum_{j';\, j' \neq j}^{n_{\Phi,j}} \left( -\Phi_{Y_j \to Y_{j'}}^{\lambda_{irr}} P_{a_{Y_j}}^{\lambda_{irr}}(t) + \Phi_{Y_{j'} \to Y_j}^{\lambda_{irr}} P_{a_{Y_{j'}}}^{\lambda_{irr}}(t) \right) \right)$$
$$+ \sum_{j';\, j' \neq j}^{n_{\Delta,j}} \left( -k_{Y_j \to Y_{j'}}^{\Delta} C_{Y_j}^{Lp,\Delta\lambda,T}(t) + k_{Y_{j'} \to Y_j}^{\Delta} C_{Y_{j'}}^{Lp,\Delta\lambda,T}(t) \right) \quad (4)$$

As a metric for the quantification of photokinetic behaviors, we define the theoretical initial velocity of $Y_j$ reaction, as $r_{Y_j}^{Lp,\Delta\lambda,T}(0)$, which takes the following expression for the reactant $Y_0 = X$, $r_X^{Lp,\Delta\lambda,T}(0) = r_{0,X}^{Lp,\Delta\lambda,T}$.

$$Theo: r_{0,X}^{Lp,\Delta\lambda,T} = -\sum_{\lambda_{irr}=\lambda_a}^{\lambda_b} \left( \sum_{j=1}^{n_{\Phi,j}} \left( \Phi_{X \to Y_j}^{\lambda_{irr}} P_0^{\lambda_{irr}} \left( 1 - 10^{-A_{tot}^{\lambda_{irr},T/\lambda_{irr}}(0)} \right) \right) \right)$$
$$- C_X^{Lp,\Delta\lambda,T}(0) \sum_{j=1}^{n_{\Delta,j}} k_{X \to Y_j}^{\Delta} \quad (5)$$

Eq.5 indicates that the initial reactant velocity, for a given reaction, is sensitive to the extrinsic parameters (i.e., the initial concentration, the optical path length, the medium temperature, and the radiation intensity entering the reactor which includes both the total absorbance of the medium, the sample's irradiated area and volume). It turns out that $r_{0,X}$ is an effective metric as shall be shown in subsequent sections.

*2.2. A general integrated rate-law model*

Since an analytically derived integrated rate-law is not achievable for Eq.1 (or Eq.4), we propose a general explicit formula (Eq.6) to map out the kinetic traces of photo-thermal reactions. It represents the first of its kind in photochemistry literature. It can be applied to traces of any species $Y_j$ irrespective of the photothermal mechanism governing the reaction. The general model equation merges the Φ-order character of the photochemical reaction-steps (the $log - exp\ (\omega\ Log(1 + cc\ e^{-k\ t}))$ functions, labelled by Φ [43-45]), and exponential-type first-order kinetics of the thermal reaction-steps (labelled by Δ). Such a formulation correlates to the Φ-order character obeyed by the traces of purely photochemical reactions driven by either monochromatic [43] or polychromatic light [44] when fitted by similar equations to Eq.6. Also, a qualitative examination of published experimental data [9,10,17,49] do not suggest that the overall trend of the photothermal reaction traces was very different from those obtained for purely photochemical ones, with some examples attempting a fitting of the traces to first-order kinetic equations [13].

$$C_{Y_j}^{Lp,\Delta\lambda,T}(t) = \omega_j^0 + \sum_{i=1}^{i_{\Phi,j}} \omega_{ij}^{\Phi} Log\left(1 + cc_{ij}^{\Phi} e^{-k_{ij}^{\Phi} t}\right) + \sum_{i=1}^{i_{\Delta,j}} \omega_{ij}^{\Delta} e^{-k_{ij}^{\Delta} t} \quad (6)$$

With the constant parameters of Eq.6 ($\omega_j^0$, $\omega_{ij}^{\Phi}$, $\omega_{ij}^{\Delta}$, $cc_j^{\Phi}$, $k_{ij}^{\Phi}$ and $k_{ij}^{\Delta}$) are worked out from the fitting of that equation to the concentration trace of species $Y_j$, $Log$ being the base 10 logarithm and $e$ the exponential function. The number $i_{\Phi,j}$ corresponds to the mono-Φ-order terms ($\omega\ Log(1 + cc\ e^{-k\ t})$), and $i_{\Delta,j}$ to the 1st-order terms that will be included in Eq.6 for a species $Y_j$. These will, up most be, respectively, equal to the numbers $n_{\Phi,j}$ and $n_{\Delta,j}$ of photochemical and thermal reaction-steps starting or ending at the considered species $Y_j$ ($i_{\Phi,j} \leq n_{\Phi,j}$ and $i_{\Delta,j} \leq n_{\Delta,j}$).

The rate of the reaction at time $t$ (equivalent to Eq.4) is obtained by differentiation of Eq.6 (with $ln$ being the natural logarithm).

$$Fit: r_{Y_j}^{Lp,\Delta\lambda,T}(t) = -\frac{1}{ln(10)} \sum_{i=1}^{i_{\Phi,j}} \frac{\omega_{ij}^{\Phi} cc_j^{\Phi} k_{ij}^{\Phi} e^{-k_{ij}^{\Phi} t}}{1 + cc_j^{\Phi} e^{-k_{ij}^{\Phi} t}} - \sum_{i=1}^{i_{\Delta,j}} \omega_{ij}^{\Delta} k_{ij}^{\Delta} e^{-k_{ij}^{\Delta} t} \qquad (7)$$

For quantification purposes, the reactant $Fit: r_{0,X}^{Lp,\Delta\lambda,T}$, is derived from Eq.7, as

$$Fit: r_{0,X}^{Lp,\Delta\lambda,T} = -\frac{1}{ln(10)} \sum_{i=1}^{i_{\Phi,j}} \frac{\omega_{ij}^{\Phi} cc_j^{\Phi} k_{ij}^{\Phi}}{1 + cc_j^{\Phi}} - \sum_{i=1}^{i_{\Delta,j}} \omega_{ij}^{\Delta} k_{ij}^{\Delta} \qquad (8)$$

From a practical viewpoint, it is useful to introduce equivalent explicit equations to describe the total absorbance traces of the reactive medium (Eq.9) and its initial velocity (Eq.10). The total absorbance is collected at both given observation wavelength, $\lambda_{obs}$, and observation optical path length, $l_{obs}$ (with, as stated above, $\lambda_{obs}$ and $l_{obs}$ not necessarily equal to their counterparts, $\lambda_{irr}$ and $l_{irr}$). The numbers, $i_{\Phi,A}$, of mono-$\Phi$-order terms and $i_{\Delta,A}$ of thermal terms, in Eqs.9 and 10, cannot respectively exceed the numbers $q$ and $u$ of, respectively, the total number of photochemical and thermal reaction-steps involved in the reaction mechanism ($i_{\Phi,A} \leq q$ and $i_{\Delta,A} \leq u$), but can be less than those numbers depending on the shape of the trace.

$$A_{tot}^{Lp,\Delta\lambda,T/\lambda_{obs}}(t) = A_{tot}^{Lp,\Delta\lambda,T/\lambda_{obs}}(\infty) + \sum_{i=1}^{i_{\Phi,A}} \omega_{i,A}^{\Phi} Log\left(1 + cc_A^{\Phi} e^{-k_{iA}^{\Phi} t}\right) + \sum_{i=1}^{i_{\Delta,A}} \omega_{i,A}^{\Delta} e^{-k_{i,A}^{\Delta} t} \qquad (9)$$

and

$$Fit: r_{0,A}^{Lp,\Delta\lambda,T/\lambda_{obs}} = \left(A_{tot}^{Lp,\Delta\lambda,T/\lambda_{obs}}(t)\right)'_{t=0} = -\frac{1}{ln(10)} \sum_{i=1}^{i_{\Phi,A}} \frac{\omega_{i,A}^{\Phi} cc_A^{\Phi} k_{iA}^{\Phi}}{(1 + cc_A^{\Phi}) \; ln(10)} - \sum_{i=1}^{i_{\Delta,A}} \omega_{i,A}^{\Delta} k_{i,A}^{\Delta} \qquad (10)$$

Therefore, quantification of the reactivity of photothermal processes by Eqs.9 and 10, in various reaction conditions, can readily be achieved by using the kinetic data of the total absorbance of the reactive medium obtained on a routine spectrophotometer. This relieves the experimentalist from acquiring individual species' traces, that generally might require extensive means and time. It is also interesting to mention that the application of Eqs.6 and/or 9, does not require a prior detailed knowledge of the actual photothermal mechanism governing the investigated system, meaning that many of the quantifications and comparisons can effectively be performed without elucidating all mechanistic aspects of the reaction (for instance, the full reaction mechanisms in play in most naphthopyran photochromes are yet not fully known, where preconceived and plausible mechanisms are still being proposed [9,50,51]). The quantification of a variety of mechanistic options and reaction conditions, using these tools, is undertaken in the following sections, but first we test the model equations for reliability.

*2.3. Reliability of the general model-equation*

The validity of Eqs.6 and 9 to map out the traces of photothermal reactions, is tested by RK-generated traces. The reliability of these equations is evidenced by their excellent fit of RK-data corresponding to a large variety of traces (more than 200) belonging to individual species (concentration traces) and total absorbances of reaction media, of various photothermal reactions. Furthermore, and in all cases, correlation coefficient values approaching unity (> 0.99) characterized the plots of both RK-simulated *vs.* Eqs.6 and 9 data, with relatively low values for both the sums-of-squared errors (*SSE* as low as 10⁻²⁰), and root-mean-square errors (*RMSE* as low as 10⁻¹³).

An example of reaction parameters and the good fit of the traces is shown in Fig.1 for the tetramolecular $XY_3(6\Phi, 5k)$ reaction involving 6 photochemical and 5 thermal reaction-steps. The polychromatic UV light (dashed line in Fig.1a) is absorbed by the four species with an *OSIA* of 60 nm (from 310 to 370 nm).

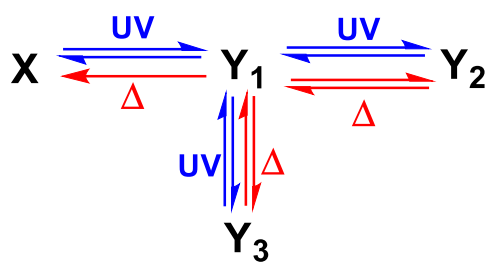

**Scheme 1:** Mechanism of $XY_3(6\Phi, 5k)$ photothermal reaction, showing photochemical (blue) and thermal (red) reaction-steps (a similar mechanism was proposed for naphthopyrans photochromic reactions [51]).

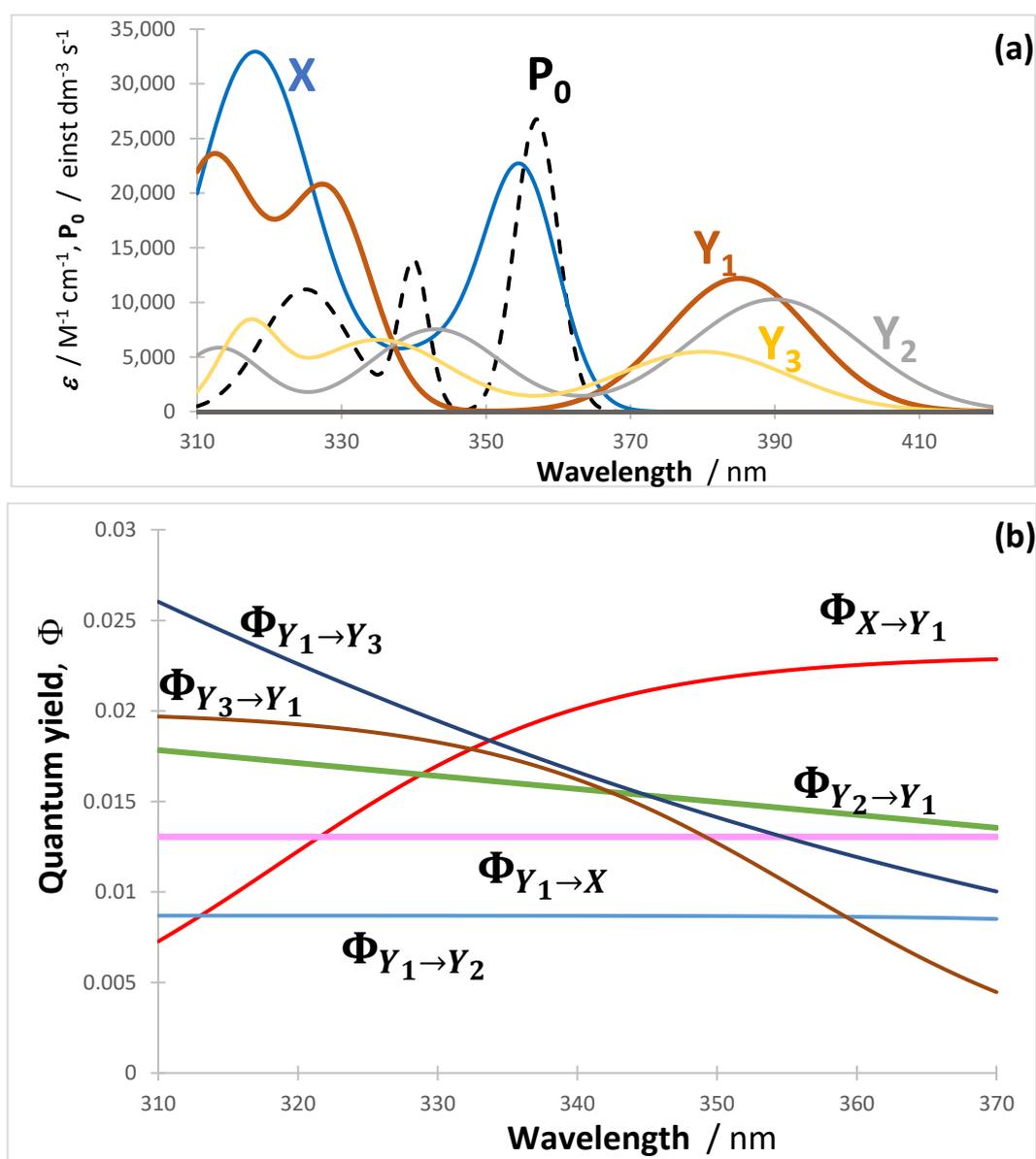

**Figure 1.** Electronic spectra ($\varepsilon_{Y_j}^{\lambda_{irr}}$), lamp profile ($P_0^{\lambda_{irr}}$), Fig.1a, and quantum yield wavelength-dependent patterns ($\Phi$), Fig.1b, of a tetramolecular reaction $XY_3(6\Phi, 5k)$, Scheme 1, proposed for *3H*-naphthopyrans [51].

The photoproducts absorb in the UV and visible, whereas the reactant only absorbs in the UV. This is typical of photochromes, such as naphthopyrans.

Most of the quantum yields are wavelength-dependent ($\Phi_{Y_1 \to X}^{\lambda_{irr}}$ and $\Phi_{Y_1 \to Y_2}^{\lambda_{irr}}$ are constant with $\lambda_{irr}$) and own independent patterns over *OSIA* (Fig.1b).

The concentration traces of the species (Fig.2) have different shapes but all reach plateau regions after long reaction time, when evolving under irradiation, which points to the reaction attaining its photothermal equilibrium or photostationary state (*pss*). The trace of the reactant is quite monotonical, whereas those of the products hint to the presence of at least two kinetic regimes. Clearly, a sole analysis of the traces cannot unravel much details on the mechanism at play (Scheme 1). The fittings Eqs.6 of the individual species' traces as well as that of the total absorbance (Eqs.9) of the reactive medium, were characterized by two mono-$\Phi$-order and one 1st-order terms. The fitting parameters of each equation (Eqs.6 and 9) were different but the fittings of the individual traces were very good ($r^2 > 0.999$).

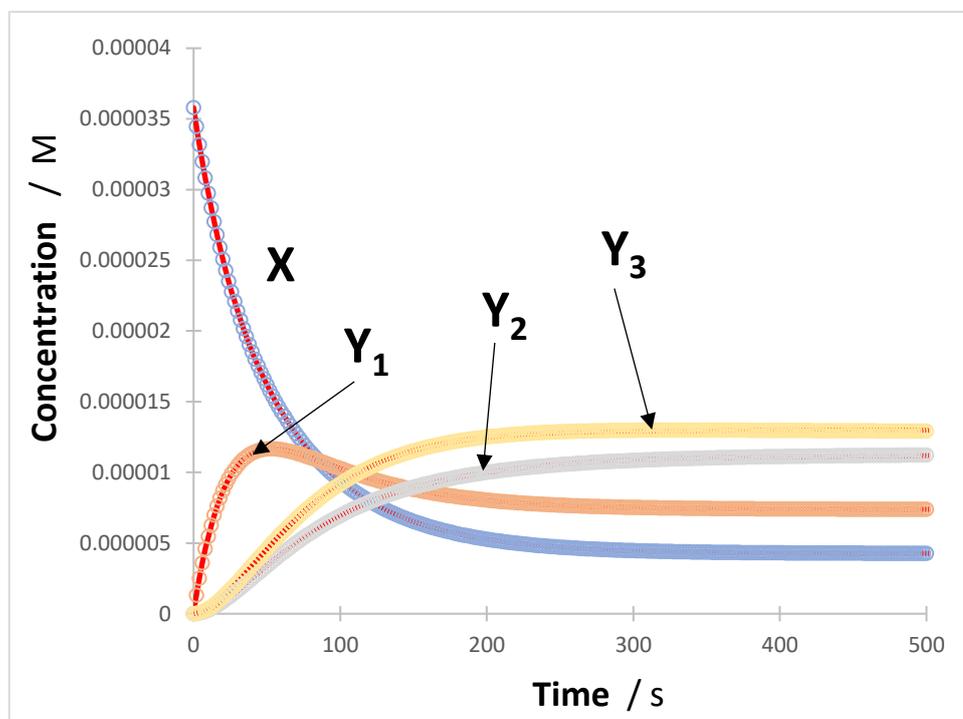

**Figure 2.** Excellent fittings of RK-calculated trace (circles) by the adequate (lines) Eq.6, for each species of the $XY_3(6\Phi, 5k)$ reaction (Scheme 1, Fig.1).

The diversity of the photomechanisms as well as that of the reaction conditions considered in the present work clearly indicates that the kinetic behavior of photothermal reactions can be regarded as a subtle combination of $\Phi$- and 1st- kinetic orders (but cannot be described by either alone, e.g., first-order kinetics). Analogous conclusions have also been reached when considering photothermal reactions under monochromatic light (in which case Eq.6 apply but the fundamental rate-laws, Eqs.1 and 4, are respectively written without the integration and the summation over $\lambda$).

Eq.6 becomes the first equation to map out photo-, thermal, and photothermal reactions in the history of photochemistry, irrespective on whether the irradiation is performed by mono- or polychromatic beams. The similarity observed on the traces obtained in the solid state [10,52,53], and those recorded for various irradiation setups [17,54] with the traces documented for the liquid state under collimated light leads to the conjecture that the scope of applicability of Eqs.6 and 9 will, most likely, extend to the kinetics of photothermal reactions that are subjected to uncollimated light. If this conjecture is true, then Eqs. 6 and 9 will be valuable not only to lab experiments but also to data collected on real-life, engineering, and industrial irradiation setups.

In addition to the good fitting of the traces, the correlation of the initial rates can further complement the proof of reliability of Eqs.6 and 9. In this context, it is necessary to confirm that $Fit: r_{0Y_j}^{Lp,\Delta\lambda,T} = Theo: r_{0Y_j}^{Lp,\Delta\lambda,T} = RK: r_{0Y_j}^{Lp,\Delta\lambda,T}$, as well as $Fit: r_{0A}^{Lp,\Delta\lambda,T} = Theo: r_{0A}^{Lp,\Delta\lambda,T}$, in all situations. With $RK: r_{0Y_j}^{Lp,\Delta\lambda,T}$ being the RK-calculated initial velocity, $Theo: r_{0Y_j}^{Lp,\Delta\lambda,T}$ is given by Eq.5, and $Theo: r_{0A}^{Lp,\Delta\lambda,T}$ is expressed by Eq.11, where $i_X$ is the set of species with non-zero concentrations at $t = 0$.

$$Theo: r_{0,A}^{Lp,\Delta\lambda,T} = l_{obs} \sum_{i=0}^{i_X} \varepsilon_{Y_j}^{\lambda_{obs}} Theo: r_{0,Y_j}^{Lp,\Delta\lambda,T} \tag{11}$$

The linear correlation between the theoretical initial velocity plotted against the other two (Fig.3), attests of both the good correspondence of these quantities and the reliability of the general model equations.

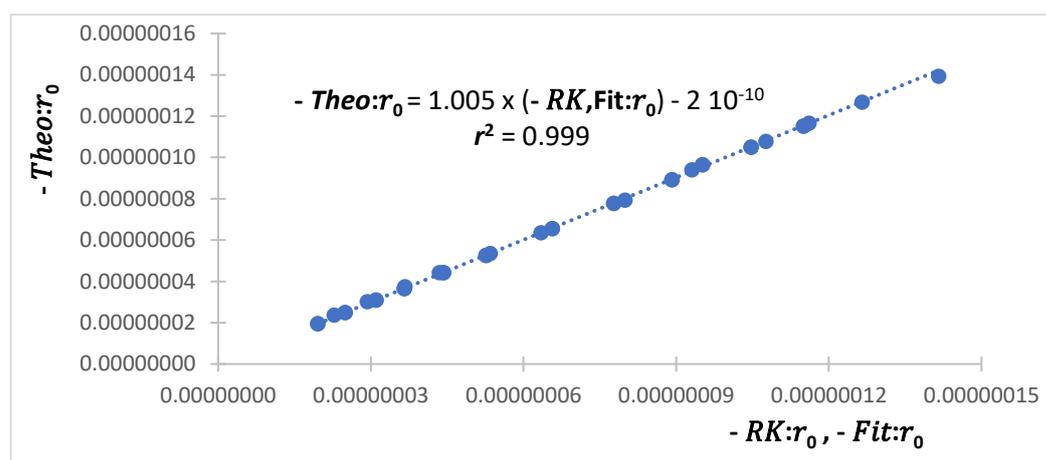

**Figure 3.** An excellent linear relationship is found between the values of $Theo: r_0^{Lp,\Delta\lambda,T}$ ($Theo: r_{0,X}^{Lp,\Delta\lambda,T}$ and $Theo: r_{0,A}^{Lp,\Delta\lambda,T}$) against the respective $RK: r_0^{Lp,\Delta\lambda,T}$ and $Fit: r_0^{Lp,\Delta\lambda,T}$ ($r_{0,X}^{Lp,\Delta\lambda,T}$, $r_{0,Y_1}^{Lp,\Delta\lambda,T}$ and $r_{0,A}^{Lp,\Delta\lambda,T}$). ($r_{0,A}^{Lp,\Delta\lambda,T}$, which have generally much larger values than $r_{0,X}^{Lp,\Delta\lambda,T}$, where scaled down by an adequate multiplicative factor). The data reported here belong to different reactive systems and experimental conditions.

Despite the good performance of the explicit equations of the model, it is useful to draw attention to two aspects that might raise confusion when applying these equations. On one hand, they don't directly allow the elucidation of the mechanism in play for the studied system. This stem from the occurrence of a distinguishability problem [43,44,49]. This is evident from the uniqueness of Eqs.6 and 9 formulations irrespective of the reaction mechanism in play. In other words, a trace (or a set of species' traces) of a given reaction sequence, can be fitted by a set of Eqs.6 corresponding to different reactions that are governed by a different mechanism, and the latter might be proposed for the reaction at hand (or conversely, traces of different reactions might well be fitted by the same equation). Hence, the elucidation of the mechanism requires the classical means of separation and identification of the products and cannot, thus far, be achieved solely by analysis of the photokinetic traces and their fitting equations. On the other hand, for a given reaction mechanism, a single trace of an individual species can well be fitted by one Eq.6 format but can have several sets of different fittings parameters [55]. The number of such diverse fitting parameters' sets can be quite large, underlining the presence of an identifiability issue [43,44,49]. In these conditions, the knowledge of the "*true*" set of values of the kinetic parameters describing the physical system under study, is impossible to single out. As a consequence, most often, the absolute values of the fitting parameters, such as the rate constants $k_{ij}^{\Phi}$, cannot be used for quantification. In this context, the only exception being the initial velocity of the reaction (including that of the reactant starting off the photothermal reaction, or the species present at the beginning of a pure thermal reaction) whose value is insensitive to the variation of the parameters' sets due to the identifiability issue. This criterion is at the basis of selecting $r_0$ as a metric.

Therefore, one can conclude that a good fit of the traces by Eqs.6 and 9 is not a proof for the veracity of neither a preconceived or proposed mechanism of the reaction nor of the fitting parameters obtained. Unfortunately, only a few developed elucidation methods are able to solve such distinguishability/identifiability issues [43,49], (with the great majority of reaction cases remain, thus far, unsolvable).

However, the validation of the explicit equations (Eqs.6 and 9) by their conformation with RK-data, promote these equations to be considered as explicit integrated rate-laws and should be treated in a similar way to any analytically derived integrated rate-law when applied to experimental data.

*2.4. Solving for the intrinsic parameters*

One of the powerful impacts of applying Eqs.6 and 9 relates to solving for the values of the intrinsic parameters (e.g., the absorption coefficients and the quantum yields) of the species involved in the reaction mechanism, as well as for the thermal rate-constants. A performance that requires three conditions. The working out of these values needs the usage of a monochromatic irradiation of the reaction. For clarification, only experimental set ups that encompass a monochromator can deliver monochromatic beams (if a monochromator is missing from the irradiation setup, then the light for irradiation is necessarily polychromatic including that produced by filtered lamp light or LEDs [43]). Also, to proceed with the solving procedure the knowledge of the complete photothermal reaction-mechanism is mandatory (preconceived mechanisms might lead to plausible values of the unknowns but these might not the true values of the chemical system at hand). Finally, both the traces of the species concentrations and that of the total absorbance of the reactive medium, must be available. In the procedure, the extrinsic parameters (experimental values of $P_0^{\lambda_{irr}}$, $l_{irr}$, $l_{obs}$, and $C_X^{Lp,\Delta\lambda,T}(0)$) are considered known. Also, the fitting equations of the traces are available. The values of $A_{tot}^{\lambda_{irr}/\lambda_{irr},T}(t)$ monitored across the optical path length of irradiation $l_{irr}$ (Eq.3), are worked out from those of the experimentally measured $A_{tot}^{\lambda_{irr}/\lambda_{obs},T}(t)$ (monitored across $l_{obs}$) as, for instance, $A_{tot}^{\lambda_{irr}/\lambda_{irr},T}(0) = A_{tot}^{\lambda_{irr}/\lambda_{obs},T}(0) \times l_{irr}/l_{obs}$.

The elucidation procedure, solving for the true set of intrinsic parameters of the reaction extends that previously set out for photoreactions [43]. The first step is to select a series of $n_t$ time intervals ($n_{sp} \leq n_t \leq 1$) within the reaction time domain. The concentrations of the species and the total absorbances, at given time-intervals, are calculated from the corresponding fitting equations. Hence, a total set of $n_t \times (n_{sp} + 1)$ data values is recorded ($n_{sp}$ concentrations and a total absorbance for each time interval of the series $n_t$). The values of $A_{tot}^{\lambda_{irr}/\lambda_{irr},T,q}(t)$ serve also the determination of the values of $P_{aY_j}^{\lambda_{irr}}(t)$ (according to Eq.2).

In a second step of the elucidation procedure, using the values of $A_{tot}^{\lambda_{irr}/\lambda_{irr},T,q}(t)$ and $C_{Y_j}^{Lp,\Delta\lambda,T,q}(t)$ obtained for the $n_t$ intervals, a series of $n_{sp}$ linearly-independent equations (Eqs.3) are established for the total absorbance at each time interval. These are linear equations, depending only on the absorption coefficients of the $n_{sp}$ species at $\lambda_{irr}$. Solving this system of linear equations leads to the determination of the absolute values of $\varepsilon_{Y_j}^{\lambda_{irr}}$ of the species.

The next stage of the elucidation consists in collecting the traces corresponding to the thermal process of the reaction evolving in the dark (by switching off the light source after irradiation, and letting the reaction to proceed thermally). These traces are fitted with Eqs.6 and 9 whose coefficients $\omega_{ij}^{\Phi}$ and $\omega_{i,A}^{\Phi}$ are equal to zero, and the corresponding $C_{Y_j}^{\Delta}(t)$ and $A_{tot}^{\Delta,\lambda_{obs}}(t)$ fitting equations are recorded for temperature $T$. The fitting equations then deliver the actual values of $C_{Y_j}^{\Delta}(t)$ and $A_{tot}^{\Delta,\lambda_{obs}}(t)$ at the specific series of $n_t$ time intervals. Let $n_\Delta$ be the number of thermal reaction-steps involved in the overall reaction mechanism. In the case $n_\Delta \leq n_t \times n_{sp}$ the traces of the species will be sufficient to solve for the set of $k_{ij}^{\Delta}$. If not, it is required to top up $n_t$ by additional time intervals, so that it becomes $n_{t'} > n_t$, and the extra $C_{Y_j}^{\Delta}(t')$ and $A_{tot}^{\Delta,\lambda_{obs}}(t')$ values are added to the set, in such a way that the number of equations is equal to $n_k = n_{t'}$. Then the values of $C_{Y_j}^{\Delta}(t)$ (and $C_{Y_j}^{\Delta}(t')$) are introduced to the corresponding rate-law equations $r_{Y_j}^{\Delta}(t)$ (Eq.12). With the values of $r_{Y_j}^{\Delta}(t)$ can be worked out from the differentiation of the fitting equations as the rate-laws (Eq.4). Hence, Eqs.12 are linear equations of $k_{Y_j \to Y_{j'}}^{\Delta}$ and $k_{Y_{j'} \to Y_j}^{\Delta}$ parameters.

$$Theo: r_{Y_j}^{\Delta}(t) = \sum_{j';\, j' \neq j}^{n_{\Delta,j}} \left( - k_{Y_j \to Y_{j'}}^{\Delta} C_{Y_j}^{\Delta}(t) + k_{Y_{j'} \to Y_j}^{\Delta} C_{Y_{j'}}^{\Delta}(t) \right) \qquad (12)$$

The system of $n_\Delta$ linearly-independent (Eqs.12) equations are solved for the $n_\Delta$ unknown $k_{Y_j \to Y_{j'}}^{\Delta}$ and $k_{Y_{j'} \to Y_j}^{\Delta}$ thermal rate-constants.

One of the alternative ways to solve for the absorption coefficients' values is to consider $n_{sp}$ equations at which the values of the concentrations and total absorbances of the thermal reaction are determined. These values are introduced in the equations of $A_{tot}^{\Delta/\lambda_{obs}}(t)$ collected from the fitting equations (as given by Eq.13 (which is equivalent to Eq.3, holding the concentrations $C_{Y_j}^{\Delta}(t)$ instead of $C_{Y_j}^{Lp,\Delta\lambda,T}(t)$).

$$A_{tot}^{\Delta,\lambda_{obs}}(t) \frac{l_{irr}}{l_{obs}} = \sum_{j=0}^{n_{sp}} A_{Y_j}^{\Delta,\lambda_{obs}}(t) \frac{l_{irr}}{l_{obs}} = \sum_{j=0}^{n_{sp}} \varepsilon_{Y_j}^{\lambda_{obs}} l_{irr} C_{Y_j}^{\Delta}(t) \qquad (13)$$

Once $k_{Y_j \to Y_{j'}}^{\Delta}$ and $\varepsilon_{Y_j}^{\lambda_{irr}}$ are knowns, we can proceed to solve for the quantum yields. These are obtained by solving the system of Eqs.4 (considered at a single time interval if the sought number of the quantum yields is $n_\Phi \leq n_{sp}$, otherwise additional Eqs.4 are measured at a sufficient number of time intervals to reach equality between the numbers of equations and $n_\Phi$). The numerical values of $r_{Y_j}^{Lp,\Delta\lambda,T}(t)$ are obtained from Eq.8, and those of the remaining coefficients in Eq.4 (except $\Phi$s) have been defined above. The system of linear Eqs.4, which now hold the quantum yields as parameters, is solved for these unknowns.

An application example of this procedure can be found in the supplementary materials for the case of a trimolecular $XY_2(3\Phi, 2k)$ photothermal reaction that has been proposed in a previous study for naphthopyrans [56,57].

In terms of a further quantification of the thermal reaction, we investigate the effect of varying the temperature of the medium on the initial-rate values of the thermal reactions (either the reactant concentration trace, $r_{0,X}^T$ or the total absorbance trace, $r_{0,A}^T$). As the effect of temperature is practically conveyed by a variation of the rate-constants of the thermal reaction-steps, $k^{\Delta}$, the simulations were caried out by feeding the RK-calculation with different values of $k^{\Delta}$ of the individual thermal reaction-steps of the reaction but ensuring that the proportionality of their respective values is maintained (i.e., the ratio of a given $k^\Delta$ values in different reactions remains constant, since a change of temperature induces a proportional variation of the reaction-steps' $k^\Delta$). From there, the temperature is worked out by multiplying each element of the set of $k^\Delta$ values of a given thermal reaction-step by an arbitrary negative constant.

We observe empirical linear relationships between the natural logarithm of the initial rates and the reverse of the temperature values (the initial rates are worked out from the reactant and total absorbance traces), as given by Eqs.14 and 15 (see supplementary information). Also, it was observed that the numerical values obtained for the slops of the two above lines are very close for a given system (less than 5% difference between $a_1$ and $a_2$, as in Eqs.14 and 15). Despite these relationships were obtained for reactions with thermally stable reactant that is regenerated by one photoproduct (e.g., $Y_1$), there is still no theoretical interpretation of these empirical formulae. Though, one may remark their resemblance with the Arrhenius formula.

$$-ln\big(Fit{:}r_{0,X}^T\big) = a_1 \frac{1}{T} + b_1 \tag{14}$$

$$-ln\big(Fit{:}r_{0,A}^T\big) = -ln\left(l_{obs} \sum_{i=0}^{i_X} \varepsilon_{Y_j}^{\lambda_{obs}} Fit{:}r_{0,Y_j}^T\right) = a_2 \frac{1}{T} + b_2 \tag{15}$$

*2.5. A tricky balance between photo- and thermal reaction-steps*

The dynamics of photothermal reactions is defined by an acute balance between the contributions of photo- and thermal reactions to the overall reaction kinetics. The latter is diversely affected when either or both the irradiation conditions and the temperature of the medium are changed. Working at relatively low temperature or relative high radiation intensity will advantage the photochemical processes, whereas reverse conditions will favor the thermal processes. At a given temperature, the 1st-order thermal reactions are insensitive to changes in the experimental conditions such as initial concentration, irradiation path length, irradiated volume…etc, while such changes may considerably affect the photochemical processes that depend on all absorbing species in the medium. Such effects will be evidenced by a change of the trace pattern and/or by the concentration values reached at the end of the reaction. A fact indicating that Eq.9 (and Eq.6) becomes very specific as soon as the conditions of the reaction are defined. Such an observation imposes the recommendation that all the reaction conditions are clearly reported in the kinetic study at hand. Also, it raises caution whenever a comparison of traces or comparison between deduced kinetic parameters are envisaged. For instance, the initial velocity is invariant with the medium temperature when the reactant is thermally inert.

*2.6. Quantification of the initial reactant concentration effect*

It has been proven that an increase of the reactant initial concentration for the case of photoreactions, induces a slowdown of reactivity [43,44]. A similar observation is also valid for the photoprocesses operating within photothermal reactions. The increase of $C_X^{Lp,\Delta\lambda,T}(0)$ results into a reduction of the photokinetic factor (Eq.2), inducing a reduction in the total amount of light absorbed by the species, which translates into a slowdown of the reaction. When the quantification is based on the negative initial velocity of the reactant, this parameter value will also decrease with increasing $C_X^{Lp,\Delta\lambda,T}(0)$, as the factors $\left(1 - 10^{-A_{tot}^{\lambda_{irr}}(0)}\right)$ in Eq.5, increase. This interpretation assumes *de facto* that the Beer-Lambert law applies, i.e., the initial concentrations concerned must fall within the linearity range of the reactant calibration graph. However, because of the presence of thermal processes in the reaction, such a photostabilisation might not be clearly noticeable for the overall reaction. Hence, for some mechanisms the slowing down of the reaction is more evident than for others. An example of the reduction of the initial velocity as resulting from an increase of the initial reactant concentration, is illustrated in Fig.4 for a $XY_2(4\Phi, 2k)$ reaction.

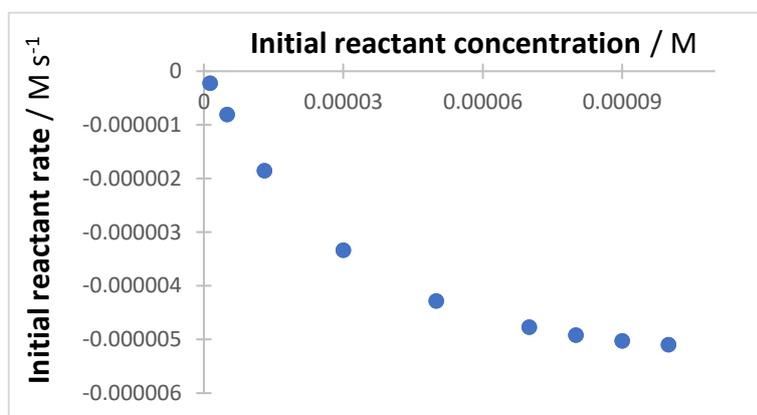

**Figure 4.** Evolution towards more negative values of the initial rate ($r_{0,X}^{Lp,\Delta\lambda,T}$) when the initial concentration ($C_X^{Lp,\Delta\lambda,T}(0)$) increases for an $XY_2(4\Phi, 2k)$ photothermal reaction.

These predictions are corroborated by similar results observed experimentally for itraconazole solutions of different concentrations exposed to polychromatic light, where the drug's degradation rate decreased as its initial concentration increased [58], (except that, in this study, the kinetic data of the reactant were fitted to first-order equations).

*2.7. Impact of spectator molecules*

The presence of photochemically, chemically and thermally stable molecule(s) in the medium, the so-called spectator molecules ($SPM$), does not in principle affect the kinetics of the thermal reaction-steps of a photothermal reaction. However, they will have an impact on the photochemical reaction-steps, provided that their absorption spectrum overlaps the $OSIA$ of the considered lamp/reaction system. In this case, the $SPMs$ absorbances (supposing there are $n_{SPM}$) make part of the total absorbance of the medium (Eq.16). Practically, this amounts to a light-shield of the reactive system, as $SPMs$ compete with the reactive species for the available incident light. An increase of $A_{SPM}^{\lambda_{irr}}$ causes both the initial velocity and the photokinetic factor to have smaller values, so much so that the kinetics of the photoreaction-steps slows down with an increase of the spectators' concentrations. In a limit case, the photochemical-steps of the reaction can virtually be stopped when a relatively high concentration of $SPM$ is used. The latter situation has been widely exploited in a qualitative manner for chromatic orthogonality under polychromatic light [59]. The interpretation offered above as well as the use of numerical integration of Eq.6 represent a fundamental explanation and a useful predictive tool for such and similar applications when the reactive materials and media are subjected to either monochromatic or polychromatic irradiations.

$$A_{tot}^{\lambda_{irr},T/\lambda_{irr}}(t) = \sum_{j=0}^{n_{sp}} A_{Y_j}^{\lambda_{irr},T/\lambda_{irr}}(t) + \sum_{j=1}^{n_{SPM}} A_{n_{SPM}}^{\lambda_{irr},T} \qquad (16)$$

The effect due the presence of $SPMs$ has been proven experimentally in our team [60,61] for reactions under monochromatic light, and RK-simulated for polychromatic light driven photoreactions [44]. The prediction laid out above have been confirmed for several photothermal reactions. For illustration, the variation of the initial velocity was monitored for an $XY_1(2\Phi, 1k)$ photothermal reaction similar to that proposed for spiropyrans [31], in the presence of increasing concentration of $SPM$, as reported in Fig.5. The resulting reduction of $r_{0,X}$ (Fig.6) is concomitant with a slowdown of the photoreactions within the photothermal process and an increased contribution of the thermal reaction-steps to the overall reaction kinetics.

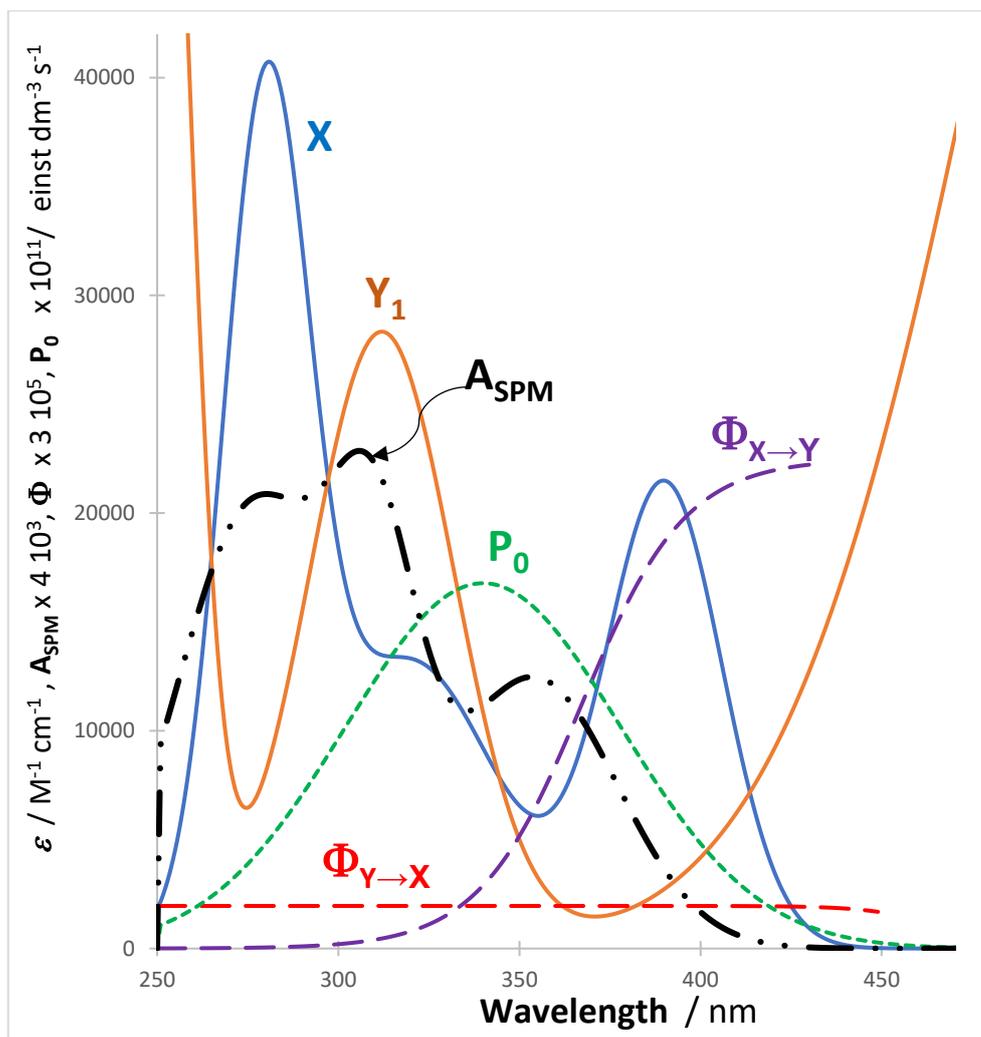

**Figure 5.** Electronic absorption of reactant (plain blue line), photoproduct (plain orange line), patterns of the quantum yields of $\Phi_{X \to Y_1}^{\lambda_{irr}}$ ($x\ 3\ 10^5$, long-dashed purple line), and $\Phi_{Y_1 \to X}^{\lambda_{irr}}$ ($x\ 3\ 10^5$, long-dashed red line), the absorption spectrum of $SPM$ ($x\ 4\ 10^3$, dotted-dashed black line), and lamp profile ($x\ 10^{11}$, dashed green line), of a photochromic $XY_1(2\Phi, 1k)$ reaction ($k_{Y_1 \to X} = 0.0028\ s^{-1}$ and $C_X(0) = 4\ 10^{-5}\ M$).

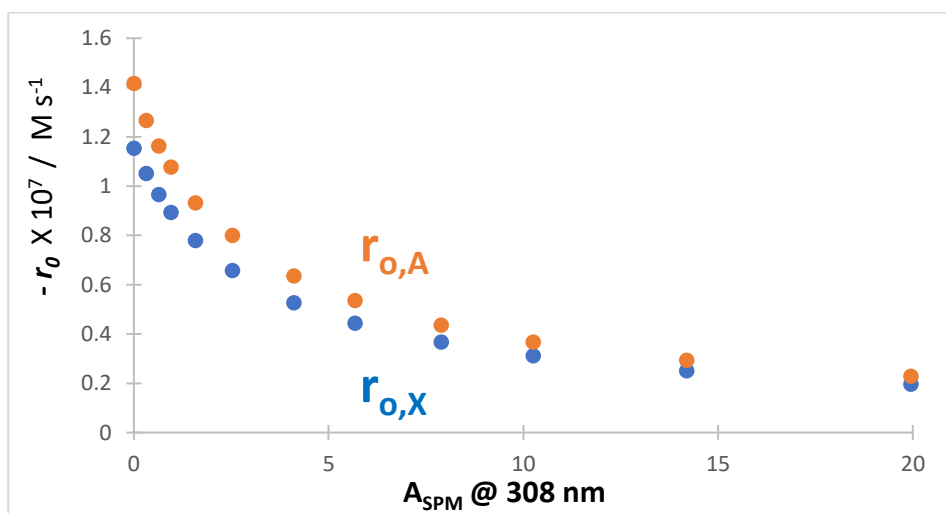

**Figure 6.** Reduction of the initial reactant velocity with increasing $SPM$ concentration represented here by the values of absorbance at 283 nm on the $SPM$ spectrum.

The effect of *SPM*s and its quantification need to be taken intoaccount for real formulations that might involve absorbing species other than those of the reaction of interest, such as in pharmaceutical formulations, environmental samples, synthesis media, biological molecules…etc.

*2.8. Kinactinometry*

As a matter of fact, there are no examples in the literature of organic photothermal reactions used as actinometers. The notorious example of the photothermal reversible E/Z-azobenzenes are generally considered for actinometry only when the reverse thermal reaction is negligible (e.g., the photoreversible reaction is studied at relatively low temperature) [62].

Photothermal chromic materials have not yet been proposed for actinometry, despite the advantages they offer relative to reversibility, a quite large spectral window, and a variety of molecular structures. The occurrence of thermal reactions in the ferrioxalate actinometer [63] are rather considered as a limitation [64] despite the popularity of that actinometer (a few of its possible limitations have also been previously underlined [44]). As far as we are aware [64,65], the development of an efficient actinometric method for $XY_v(q\Phi, uk)$ photothermal reactions is unprecedented in the literature.

Kinactinometry is kinetic data based actinometry (aimed at determining the incident light intensity, i.e., $P_0$). Kinactinometry for photothermal reactions correlates the initial reaction velocity (as a metric) to the number of photons either entering the reactor as a whole or the number of photons absorbed by the reactant of the actinometric reactive system. These two numbers are not necessarily equal for all reaction conditions. As such, this kinactinometry relays a property of either the reactant and does not depend on the mechanism or the rest of species involved in the reaction, i.e., the metric is $r_{0,X}^{Lp,\Delta\lambda,T}$, or it conveys the features of both the reactant and its immediate photoproducts when the metric is $r_{0,A}^{Lp,\Delta\lambda,T}$. The specificity of the former allows control of the wavelength domain of actinometry as it corresponds to the *OSIA* of the actinometer with the lamp profile (irrespective of the absorption spectra of the products, which might be unknown to the experimentalist) but requires the acquisition of the reactant trace. The latter might be more convenient and easier to implement but the metric ($r_{0,A}^{Lp,\Delta\lambda,T}$) encompasses the absorption spectra of the reactant and its immediate products causing the precise knowledge of the wavelength domain of the actinometer to be reliant on the knowledge the electronic spectra of these products (if the latter are not available, an uncertainty on the actinometric wavelength domain will persist).

We know that the kinetic traces recorded experimentally or RK-calculated correspond solely to those photons entering the reactor that are absorbed by the reactive system (e.g., the actinometer). Hence, fitting the traces to Eq.6 reflects the true physical system described by Eq.4. Furthermore, Eq.6 only informs about the absorbed photons that induce change in the reactive system, for reactivity is only observed when the radiation intensity, the absorption coefficient as well as the quantum yield, have all non-zero values at $\lambda_{irr}$ (i.e., according to Eq.4, absorption of light is a necessary but not a sufficient condition).

In accordance with Eq.5, a linear proportionality (of a typical formula, $r_0 = grad\ P_0 + intercept$) is expected for the variation of the initial reaction velocity ($r_0$) with incident radiation intensity ($P_0^{\lambda_{irr}}$), where *grad* and *intercept* are, respectively, the slop of the line and its intercept.

The RK-calculated traces of several photothermal systems were fitted to Eq.6 (or Eq.9) and the corresponding initial velocities were determined and plotted against the incident radiation intensity. Linear relationships were obtained in all cases. The lines were characterized by correlation coefficients close to unity, and intercepts either close to zero for thermally inert reactants, and different of zero when the reactant is thermally active (Fig.6). These correlations were established for the reactant (including *theo*: (Eq.5), *Fit*: (Eq.8), and $RK: r_{0,X}^{Lp,\Delta\lambda,T}(t)$), and the total absorbance of the medium initial velocities (including *theo*: (Eq.11), *Fit*: (Eq.10), and $RK: r_{0,A}^{Lp,\Delta\lambda,T}(t)$) versus the radiation intensity (Fig.7). The gradient of the linear plot serves the subsequent determination of an unknown radiation intensity of the same lamp to which the actinometer is subjected (as $P_{0,unk} = (Fit: r_{0,unk} - intercept)/grad$). The radiation intensity ($P_0^{\lambda_{irr}}$) is measured at the wavelength of irradiation when monochromatic light is used, whereas a sum of radiation intensities at the individual wavelengths of *OSIA*, combined in the total incident light intensity ($P_{0,tot}^{Lp,\Delta\lambda,T} = \sum P_0^{\lambda_{irr}}$), is obtained when polychromatic light is employed.

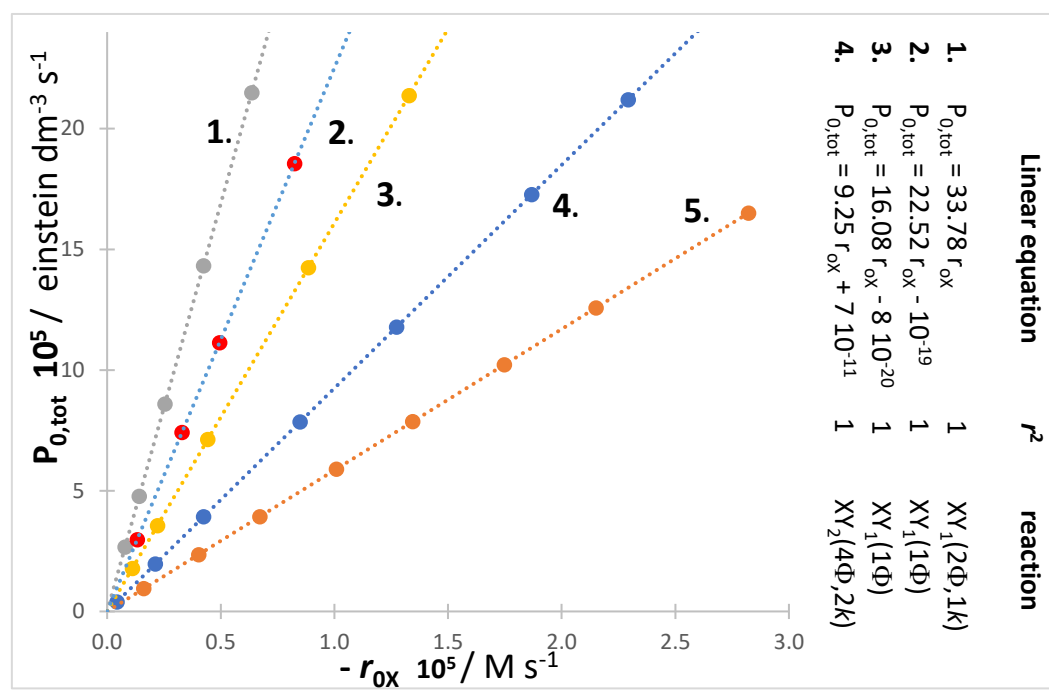

**Figure 7.** Examples of linear relationships of the initial velocity *vs.* incident radiation intensity for various photothermal reactions proposed in the literature.

Developing kinactinometric methods on the basis of the initial velocity and Eqs.6 or 9, relieves the experimentalist from the conditional knowledge of the detailed photothermal mechanism in play and/or a none wavelength-dependency of the quantum yields, as usually stated for actinometry [65]. The usage of the total absorbance of the medium further relieves from the necessity of acquiring the traces of the reactant and/or the individual species involved in the reaction. The above method represents the first of its kind in the photochemistry literature for the application of photothermal reactions to efficient, reliable and easy to implement actinometry.

Despite the efficacy of the kinactinometric method, there is still some general observation on actinometry that would be relevant to discuss. If there are no issues for actinometry of monochromatic light, it is not clear which total incident light needs to be used for the standardization of the actinometer when subjected to polychromatic light. In the above procedure, the wavelength section covering the *OSIA* was considered for the determination of $P_{0,tot}^{Lp,\Delta\lambda,T}$. However, we have seen that the photons responsible of reactivity are those that coincide with an absorption of the system that has a non-zero quantum yield. The data on absorption are provided by *OSIA* but the information of the quantum yield variation with wavelength over that *OSIA*, is most often not available. Also, the method (physical or chemical actinometry) used for the determination of $P_{0,tot}^{Lp,\Delta\lambda,T}$ can have an impact on the standardization of the actinometer. Generally speaking, among the set of photons delivered by the lamp to the reactor, there might be some whose wavelengths lay outside *OSIA*. The latter group effectively enter the reactor but do not contribute at all, to the reaction (because for such wavelengths, $P_0^{\lambda_{irr}} \neq 0$ but $\varepsilon^{\lambda_{irr}} = 0$ or $\Phi^{\lambda_{irr}} = 0$). The way of measuring the incident number of photons may well not distinguish between these two types and hence leading to confusion in the actinometric measurement.

For instance, if the number of photons of an unfiltered wide-range lamp, as classically used, is evaluated by the ferrioxalate actinometer, then all the photons emitted by that lamp with a wavelength under 500 nm will, presumably, all be detected and accounted for in the final number of photons delivered to the reactor. But in general, the actinometer to be developed (or the reaction studied) has a narrower absorption spectrum (e.g., 200 – 380 nm) than that of the ferrioxalate actinometer (which roughly spans the 200-500 nm section). Conversely, using a physical actinometer that is limited to *OSIA*, avoids counting the extra photons falling outside *OSIA* but unfortunately does not prevent counting those where $\Phi^{\lambda_{irr}} = 0$ if, for the considered system, the quantum yield pattern with wavelength is not precisely known. Here, one should bear in mind that linear relationships will still be obtained for $r_0 = f(P_0)$ irrespective on whether the total number of photons corresponds to only those inducing the reaction, or includes those which do not (e.g., laying outside *OSIA*), as was mathematically derived and proven for photoreactions' kinactinometry [44]. In any case, measuring kinactinometrically the total number of photons that is delivered by the lamp, is crucial to determining the photonic yield as we shall see in the next section.

Accordingly, an actinometric evaluation of the incident number of photons must specify whether the determined number corresponds to what is absorbed only, or encompasses a larger number emitted by the lamp (including filtered and unfiltered lamps, and LEDs). Of course, if the span of wavelengths covered by the lamp (e.g., LED lights) is narrower than the absorption spectral section of the actinometer or the reactive system, then the incident number of photons measured by, for instance, the ferrioxalate actinometer, is valid for *OSIA*-actinometry.

Another confusion about actinometry relates to the assumption that the number of photons determined by the actinometer are absorbed by the investigated system and all induce reactivity. In practice, the *OSIA*s of actinometer and investigated system are not always or necessarily equal. But if they are, the quantum yield profiles of the photochemical-steps in the investigated system might be very different from those exhibited by the actinometer. In any case, it was stated above that Eq.4 (and by inference, Eq.6 as well) is specific to both the system under study and the experimental conditions selected for the reaction. Hence, knowing the number of photons by kinactinometry does not make Eq.4 of the actinometer and that of the investigated system, the same. In summary, one is forced to admit that, for polychromatic light, the best actinometer would arguably be the investigated system itself where the quantification is achieved by a kinactinometric method based on fitting the traces with Eq.6 (or Eq.9).

Finally, the specificity of Eq.4 to reaction and reaction conditions implies that an actinometer standardized for one lamp cannot be used to determine the number of photons delivered by another lamp. Similarly, different kinactometric features would characterize two actinometers subjected to the same lamp. These matters have been discussed at length in a previous study dealing with photochemical reaction [44].

*2.9. Photonic yield*

We have seen in *Section 2.4* that the absolute values of the quantum yield at a specific wavelength ($\lambda_{irr}$) are worked out by the procedure solving for the intrinsic parameters, when monochromatic irradiation is employed. In fact, the quantum yield can only be defined if the irradiation light is monochromatic [65].

When the photothermal reaction is exposed to polychromatic light, the photonic yield (*PY*, which is dimenssionless) is often measured. It expresses the extent of reactivity for a set of photons entering the reactor. *PY* can be defined for the reactant by the ratio of the initial velocity and the incident light intensity (Eqs.17 and 18), as has been established for photoreactions.

$$PY_{0,X}^{Lp,\Delta\lambda,T} = -\frac{r_{0,X}^{Lp,\Delta\lambda,T}}{P_0^{Lp,\Delta\lambda}} \qquad (17)$$

$$PY_{0,A}^{Lp,\Delta\lambda,T} = -\frac{r_{0,A}^{Lp,\Delta\lambda,T}}{P_0^{Lp,\Delta\lambda}} \qquad (18)$$

However, the reactant might not relay a full picture when thermal reactions are involved in the mechanism. We propose here a more inclusive definition of *PY* (Eq.19) to render account of the reactivity/intensity extent over a larger span of time, applicable for any species of the reaction. The required concentrations (or absorbances) for the definition of *PY* at the selected time intervals can precisely be defined from the corresponding fitting equations (Eqs.6 or 9). In Eq.19 the pre-fraction signs ensure that the dimensionless *PY* has positive values in all cases. This *PY* definition is analogous to the one found in the literature [65,66], except that the latter is specific to the reactant.

$$PY_{Y_j}^{Lp,\Delta\lambda,T} = \pm \frac{C_{Y_j}^{Lp,\Delta\lambda,T}(0) - C_{Y_j}^{Lp,\Delta\lambda,T}(t)}{P_0^{Lp,\Delta\lambda}\, t} \qquad (19)$$

Or by using the total absorbance data, with now $PY_A^{Lp,\Delta\lambda,T}$ is expressed in $einstein\ dm^{-3}$ units.

$$PY_A^{Lp,\Delta\lambda,T} = \pm \frac{A_{Y_j}^{Lp,\Delta\lambda,T}(0) - A_{Y_j}^{Lp,\Delta\lambda,T}(t)}{P_0^{Lp,\Delta\lambda}\, t} \qquad (20)$$

It must be kept in mind that, conversely to the quantum yield, the *PY* (Eqs.17-20) is a relative quantity and not a property of the material. It varies with reaction time and/or any factor that alter the kinetic traces, i.e., all experimental conditions including, for instance, changes occurring on the incident radiation intensity, the irradiation path lengths, or temperature (Fig.8).

To shed light on *PY* variability, let us consider its pattern with several $P_0^{Lp,\Delta\lambda}$ values on the reactant (*X*) and a product ($Y_j$) of an $XY_v(q\Phi, pk)$ reaction.

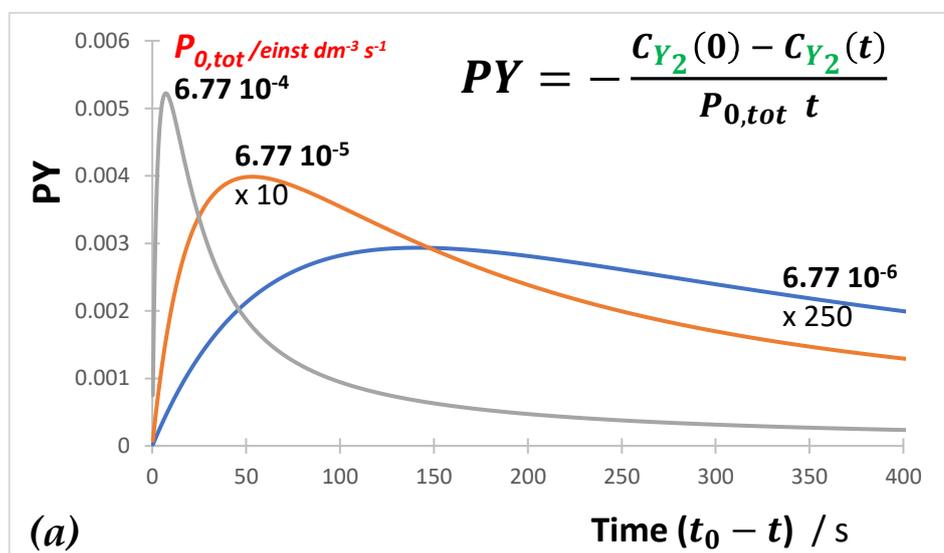

(a)

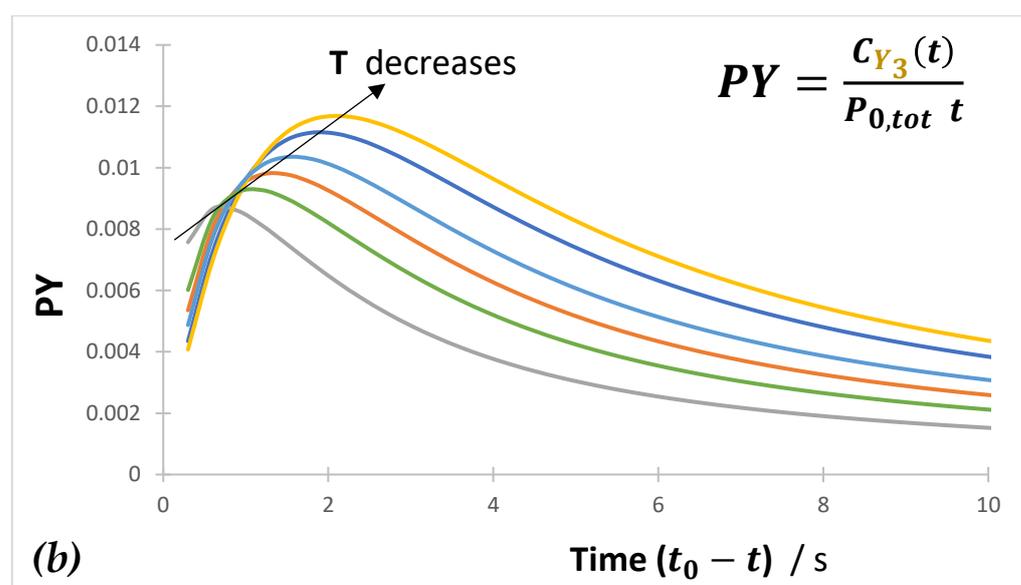

**Figure 8.** Examples of linear relationships of *PY vs.* time of intensity for $XY_2(4\Phi, 2k)$ reaction. The photonic yield is determined for a variation of either the incident light intensity (*a*) or the temperature of the medium (*a*).

The variability of $PY$ imposes caution before reaching reliable conclusions when comparing $PY$ values, because comparison means that there must be some reaction factor(s) that have changed between the systems to be compared.

A great number of photothermal mechanisms were proposed in the literature for photochromic materials. Very few of those were proven experimentally (most of these are relatively simple involving less than 4 reaction-steps), so much so that for the majority of complex reactions, potentially involving four or more reaction-steps, the mechanisms reported in the literature for photochromes are preconceived or based on experimental observations and a kinetic treatment that yields plausible values for the intrinsic and reactivity parameters. In a previous study [31,49,50,56,57], it was analytically proven that the latter criteria were far from being sufficient to represent a proof for the validity of a proposed preconceived mechanism. Indeed, a single kinetic data-set experimentally collected for a naphthpyran thermal reaction ($XY_2(nk)$), could readily be described by no less than eight different mechanisms (with $2 \leq n \leq 4$), indicative of the occurrence of a distinguishability issue). Similarly, one data series of the photothermal reaction of that naphthopyran was studied under the assumption that it obeys a previously proposed preconceived mechanism, $XY_2(3\Phi, 2k)$ [67]. In this case, it was demonstrated that several different sets of plausible values of the reaction intrinsic parameters, all delivered excellent fits of the kinetic traces by the corresponding equations, which revealed a serious identifiability problem. Because, such a degeneracy of the kinetic solution has not yet been solved, the true mechanism of naphotopyrans and many other photothermal reactions are not known with certainty. As a consequence, it is not possible to work out the absolute values of the quantum yields and the absorption coefficients (for the thermally active species) using the procedure laid out above, that solves for the intrinsic parameters, since the knowledge of the reaction mechanism was a pre-requisite. In these conditions, and from a kinetic analysis viewpoint, only the photonic yield can be determined to characterize the reactions whose mechanism is not known.

**3. Experimentals**

Numerical integration methods (*NIM*s) have usually been used in photokinetics to fit photo- and photothermal reactions' traces [2], hence justifying the usefulness and reliability of these methods in this area. Their performance in describing and predicting the kinetic behaviors of photoreactions has also been proven in previous studies [43,44]. *NIM*s are used in the present work to forecast the kinetics and generate the corresponding traces of photothermal reactions considered under various conditions.

The methods developed here are intended for reactions of the type $XY_v(q\Phi, uk)$, that represent, but not limited to, photothermal reactions encountered in photochemistry literature. These monomolecularly initiated processes include a reactant ($X = Y_0$) and $v$ products, $Y_v$, generated by photochemical and/or thermal processes. The $n_{sp}$ ($= v + 1$) species are interlinked in the overall mechanism of the photothermal reaction by "$q$" photochemical reaction-steps characterised each by an individual quantum yield ($\Phi$), and "$u$" purely thermal reaction-steps proceeding by individual rate-constants ($k^T$). The main reaction examples shown in the above sections have been proposed in the literature for many photochromes, including non-reversible $XY_1(1\Phi, 1k)$ [68], a reversible $XY_1(1\Phi, 1k)$ [69], $XY_1(2\Phi, 1k)$ [72], $XY_1(6\Phi, 5k)$ [51], $XY_2(3\Phi, 2k)$ [57], and $XY_2(4\Phi, 2k)$ [9,56].

The rate-laws for the present work, are written for a slab-shaped, vigorously and continuously stirred reactor. The reactive medium is exposed to a collimated light beam (whether monochromatic or polychromatic). The rate-equations also consider that the employed concentrations, of all species at all times, must fall within the linearity ranges of the species respective calibration graphs. It has been suggested that, in order to satisfy this condition, the total absorbance of the reactive medium at any $\lambda_{irr}$ should not exceed a value of 0.5, which might well translate into individual species absorbances belonging to their respective linearity ranges for many reactive systems [43,44].

The concentration kinetic traces of species $Y_j$ of a given photothermal reaction (i.e., the variation with time of the species concentration, $C_{Y_j}^{Lp,\Delta\lambda}(t)$), were calculated, for each species, by the fourth-order Runge-Kutta (*RK*-4 numerical integration method. The method was run, by a home-made program, on the *VBA* platform of Microsoft Excel.

The concentration traces obtained by *RK*-simulations (for $Y_j$) were fitted by the general equation (Eq.6). The total absorbance traces were fitted by Eq.9. The fitting procedure was performed by a Levenberg-Marquardt Algorithm (*LMA*), which is part of the curve fitting tool of the R2020b MatLab software. The quality of the fittings was assessed by three criteria, namely, (i) the squared correlation coefficients ($r^2$), of the linear plots constructed with the RK-generated and *LMA*-calculated data of each trace, (ii) the sum-of-squares error (SSE), and (iii) the root-mean-square deviation (*RMSD*) between the two data sets.

## 4. Conclusion

Runge-Kutta numerical integration has been proven to be an efficient tool to reliably simulate the kinetic behavior of $XY_v(q\Phi, uk)$ photothermal reactions irrespective of the number of species, photochemical and thermal reaction-steps involved in the overall process. It represents an interesting qualitative/quantitative tool to explore the kinetics of $XY_v(q\Phi, uk)$ in different reaction and irradiation conditions.

The general explicit model equation, Eq.6 (or Eq.9), has been proven to map out the kinetic traces of $XY_v(q\Phi, uk)$ with high accuracy. This model equation represents the first example of its kind that is applicable to any photothermal reaction. Its usefulness in quantifying the most ubiquitous reaction situations has been demonstrated even though the formulation of the general model equation could not be proven analytically.

The model equation has also served the development of reliable kinactinometric methodologies that allow the measurement of either the number of photons absorbed by the actinometer or those delivered by the lamp (before absorption). Such kinactinometric methods have opened a new avenue for the recruitment of photothermal reactions, in general, and photochromic materials in particular, as a class of molecules that, up to date, has not been envisaged for actinometry.

In the present work, a photonic-yield formulae have been proposed that apply either to any species in the reaction, or to the total absorbance, using in each case the specific kinetic trace and the corresponding general model equation by which it is fitted.

Overall, the results of the present study have demonstrated that quantification of the reaction and actinometry can be achieved by simply using the absorbance of the reactive medium (that can be obtained on a routine spectrophotometer) and the fitting of that trace to Eq.9 (using one of the common and commercially available softwares), without a prerequisite knowledge of the precise mechanism of the reaction, the exact number of the species involved in the reaction, the reaction species' quantum yield and/or absorption coefficients patterns with wavelength. The facility by which kinetic investigation can be carried out on $XY_v(q\Phi, uk)$ reactions, by using the general model equation, is unprecedented in the literature.

A method solving for the intrinsic parameters ($\Phi$ and $\varepsilon$) has also been described when the mechanism of the reaction is known and data of monochromatic irradiation is available.

Furthermore, it is conjectured that the general model equation (either Eq.6 or 9) would be applicable to experimental setups including any sample and radiation geometries, and/or when exposed to uncollimated polychromatic light.

A similar strategy is currently under development for the quantification of bimolecular photoreactions of the type $X + X' \to Y$.

**Supplementary Materials:** The following supporting information can be downloaded at: …..

**Author Contributions MM:** conception, design and writing, and approval of the study.

**Funding:** This work did not receive financial support.

**Institutional Review Board Statement:** Not applicable.

**Informed Consent Statement:** Not applicable.

**Data Availability Statement:** Data are contained within the article and Supplementary Materials.

**Conflicts of Interest** The author declares that the research was conducted in the absence of any commercial or financial relationships that could be construed as a potential conflict of interest.

*Glossary of symbols*

| | |
|---|---|
| $A$ | Absorbance |
| $a$ and $b$ | Coefficients of the linear relationships, Eqs.14 1nd 15 |
| $A_{tot}^{\lambda_{irr}/\lambda_{irr},T}(t)$ | Total absorbance of the reactive medium at temperature ($T$) and wavelength $\lambda_{irr}$ at time $t$ across $l_{obs}$ (also trace of medium's total absorbance). The second $\lambda_{irr}$ might be equal to $\lambda_{obs}$ |
| $A_{Y_j \text{ or } j'}^{\lambda_{irr}/\lambda_{irr},T}(t)$ | Absorbance of species $Y_j$ (or $Y_{j'}$) measured at temperature ($T$) and wavelength $\lambda_{irr}$ and $l_{irr}$ at time $t$ (also trace of $Y_j$ (or $Y_{j'}$) absorbance) |
| $A_{tot}^{Lp,\Delta\lambda,T/\lambda_{obs}}(t)$ | Total absorbance of the reactive medium at time $t$ irradiated by lamp $Lp$ and observed at $\lambda_{obs}$, where the optical path length is $l_{obs}$ (also trace of medium's total absorbance) |
| $A_{tot}^{Lp,\Delta\lambda,T/\lambda_{obs}}(\infty)$ | Total absorbance of the reactive medium at the end of the reaction ($t = \infty$) |
| $A_{n_{SPM}}^{\lambda_{irr},T}$ | Total absorbance of the spectator molecules, present in reactive medium ($0 \leq r \leq w$) |
| $A_{tot}^{\Delta,\lambda_{obs}}(t)$ | Total absorbance of the reactive medium evolving in the dark at temperature ($T$) and observed at wavelength $\lambda_{obs}$ at time $t$ across $l_{obs}$ (also trace of medium's total absorbance in the dark). |
| $C_{Y_j}(t)$ | The value of $Y_j$ concentration at time $t$ (in $M$ or $mol\ dm^{-3}$), or the kinetic traces of $Y_j$. It can be labelled with specific features, such $Lp, \Delta\lambda, T,$ or $\Delta$ (e.g., $C_{Y_j}^{\Delta}(t)$ for the experiment performed in the dark) |
| $cc_j$ , $cc_A$ | Pre-exponential factors within the mono-$\Phi$-order terms in, respectively, Eqs.6 and 9, for species $Y_j$. They are labelled by either $\Phi$ or $\Delta$ |
| $\Delta$ | label for thermal processes |
| $\Delta\lambda$ | Wavelength span of the polychromatic light impinging of the reactor, $\Delta\lambda = \lambda_b - \lambda_a$ |
| $e$ | Exponential function |
| $\varepsilon_{Y_j}^{\lambda_{irr} \, or \, \lambda_{obs}}$ | Absorption coefficient of species $Y_j$ at $\lambda_{irr}$ or $\lambda_{obs}$ (in $M^{-1} \, cm^{-1}$) |
| $j \, or \, j'$ | Index for the species $Y$ ($Y_j$), (or $Y_{j'}$, with $j' \neq j$) |
| $i_{\Phi,A}$, $i_{\Delta,A}$ | Numbers of mono-$\Phi$-order and 1st-order terms, in Eqs.9 and 10 |
| $i_{\Phi,j}$, $i_{\Delta,j}$ | Numbers of mono-$\Phi$-order and 1st-order terms, in Eq.6 |
| $i_X$ | Set of species with non-zero concentrations at $t = 0$ |
| $k$, $k_{ij}$, $k_{iA}$ | Rate-constant (in $s^{-1}$) of a reaction, or of the $i^{th}$ regime of the concentration trace of species $Y_j$, or on the total absorbance trace. It can be labelled by $\Phi$ or $\Delta$ |
| $k_{Y_j \to Y_{j'}}^{\Delta}$ | Rate-constant of thermal reaction-step $Y_j \to Y_{j'}$ ($j \neq j'$) |
| $k_{Y_{j'} \to Y_j}^{\Delta}$ | Rate-constant of thermal reaction-step $Y_{j'} \to Y_j$ ($j' \neq j$) |
| LMA | Levenberg-Marquardt Algorithm |
| $ln$ | Natural Logarithm, base $e$ |
| $Log$ | Logarithm base 10 |
| $Lp$ | The lamp used for irradiation of the sample |
| $l_{irr}$ | Optical path length of the irradiation light from the lamp inside the reactor (in $cm$) |
| $l_{obs}$ | Optical path length of the monitoring light from the spectrophotometer inside the reactor |
| $\lambda_{irr}$ | Irradiation wavelength (in $nm$, the wavelength of the light beam driving the reaction and different from the wavelengths of isosbestic points) |
| $\lambda_{obs}$ | Observation wavelength, at which the medium is monitored. |
| MatLab | Software for trace fitting |
| NIMs | Numerical integration methods |
| $v + 1$ | Total number of the $Y_j$ species in the a given photothermal reaction's photomechanism, ($0 \leq j \leq n_{sp}$), with $Y_0 = X$ for the reactant and the $v$ $Y_{j>0}$ products. |
| $n_{\Phi j}$ | The number of photoreaction-steps starting or ending at species $Y_j$ ($1 \leq n_{\Phi j} \leq q$) |
| $n_{\Delta j}$ | The number of thermal-steps starting or ending at species $Y_j$ ($1 \leq n_{\Delta j} \leq u$) |
| $n_{SPM}$ | The number of spectator molecules in the medium |
| $n_{sp}$ | Species present at time $t$ and absorbing at $\lambda_{irr}$ |
| OSIA | Wavelength span corresponding to the incident irradiation light (from the lamp) that is absorbed by the considered species |
| $PKF(t)$ | Photokinetic factor of the reaction at hand at time $t$ |
| $P_{0,tot}^{Lp,\Delta\lambda,T}$ | Total number of photons entering the reactor, $P_0^{Lp,\Delta\lambda} = \sum_{\lambda_a}^{\lambda_b} P_0^{\lambda_{irr}}$ |
| $P_0^{\lambda_{irr}}$ | Incident radiation intensity from the lamp at $\lambda_{irr}$, entering the reactor (it is a flux of photons per $s$ per the sample's irradiated surface $S_{irr}$ and volume $V_{irr}$). It is expressed in $einstein \, s^{-1} \, dm^{-3}$ |
| $P_{a_{Y_j \, or \, j'}}^{\lambda_{irr}}(t)$ | Time-dependent amount of the light absorbed by the species at the start of the considered photochemical reaction-step at time $t$ and at wavelength $\lambda_{irr}$ |
| $\Phi$ | Quantum yield of a given photoreaction-step |
| $\Phi_{Y_j \to Y_{j'}}^{\lambda_{irr}}$ | Quantum yield of photoreaction-step $Y_j \to Y_{j'}$ ($j \neq j'$) |
| $\Phi_{Y_{j'} \to Y_j}^{\lambda_{irr}}$ | Quantum yield of photoreaction-step $Y_{j'} \to Y_j$ $(j' \neq j)$ |
| $PY_{0,X \text{ or } A}^{Lp,\Delta\lambda,T}$ | Photonic yield measured on the basis of initial rates. It is dimensionless and time-independent |
| $PY_{Y_j \text{ or } A}^{Lp,\Delta\lambda,T}$ | Photonic yield measured on the basis of a species concentration or the total absorbance. It is time-dependent, dimensionless when concentration based, and expressed in $einstein\ dm^{-3}$ when it is total absorbance based |
| $q$ | Total number of the photochemical reaction-steps in the considered photothermal mechanism |
| $r^2$ | Squared correlation coefficients |
| $RK$ or $RK4$ | Fourth-order Runge-Kutta numerical integration method ($RK4$-$NIM$) |
| $RMSE$ | Root-mean-square deviation |
| $r^{Lp,\Delta\lambda}(t)$ | Contribution of the $n_{\Phi,j}$ photochemical reaction-steps related to $Y_j$ to the general species rate-law $r_{Y_j}^{Lp,\Delta\lambda,T}(t)$ |
| $r^{\Delta}(t)$ | Contribution of the $n_{\Delta j}$ photochemical reaction-steps related to $Y_j$ to the general species rate-law $r_{Y_j}^{Lp,\Delta\lambda,T}(t)$ |
| $r_{Y_j}^{\Delta}(t)$ | Rate-constant of the thermal reaction-step starting with species $Y_j$ (expressed in $M\ s^{-1}$) |
| $r_{Y_j}^{Lp,\Delta\lambda,T}(t)$ | Rate of species $Y_j$ at time $t$ (in $M\ s^{-1}$) or rate-law for species $Y_j$ |
| $r_{Y_0}^{Lp,\Delta\lambda,T}(0)$ | Initial rate of the reactant (in $M\ s^{-1}$), ($r_{Y_0}^{Lp,\Delta\lambda,T}(0) = r_X^{Lp,\Delta\lambda,T}(0) = r_{0,X}^{Lp,\Delta\lambda,T}$) |
| $r_{Y_j}^{\lambda_{irr}}(t)$ | Rate at wavelength $\lambda_{irr}$ and time $t$ (in $M\ s^{-1}$) for species $Y_j$, or rate-law for that species |
| $r_{0,Y_j}^{\lambda_{irr}}$ | Initial reaction-rate of species $Y_j$ $(0 \leq j \leq n_{sp}$, with for the reactant, $Y_0 = X)$ |
| $r_{Y_j}^{T}(t)$ | Rate at time $t$ (in $M\ s^{-1}$) for species $Y_j$, or rate-law for that species. Reaction in dark |
| $Fit: r_{0,Y_j}^{Lp,\Delta\lambda,T \text{ or } \Delta}$ | Initial reaction-rate of species $Y_j$ calculated by using trace fitting to Eq.6 and its parameters |
| $Fit: r_{0,A}^{Lp,\Delta\lambda,T \text{ or } \Delta}$ | Initial reaction-rate of species $Y_j$ calculated by using the total absorbance trace's fitting equation (Eq.9 or 12) and its parameters |
| $RK: r_{0,Y_j}^{Lp,\Delta\lambda,T \text{ or } \Delta}$ | Initial reaction-rate of species $Y_j$ calculated by RK |
| $RK: r_{0,A}^{Lp,\Delta\lambda \text{ or } \Delta}$ | Initial reaction-rate of calculated by RK for the total absorbance trace |
| $Theo: r_{0,Y_j}^{Lp,\Delta\lambda,T \text{ or } \Delta}$ | Initial reaction-rate of species $Y_j$ calculated from the rate-law equation (Eq.4) |
| $Theo: r_{0,A}^{Lp,\Delta\lambda,T \text{ or } \Delta}$ | Initial reaction-rate calculated from a linear combination, Eqs.11 |
| $sp$ | Index relative to the species |
| $SPM$ | Spectator molecule(s) which are both thermally and photochemically inert |
| $SSE$ | Sum-of-squares error |
| $t$ | Reaction time (in $s$). |
| $T$ | Temperature of the reactive medium (in $°C$). |
| $u$ | Total number of the thermal reaction-steps in the considered photothermal mechanism |
| $VBA$ | Visual Basic Applications |
| $\omega_j^0$ | Constant parameter of Eq.6 (also obtained by fitting of the data to Eq.6). |
| $\omega_{ij}, \omega_{i,A}$ | Pre-logarithmic and pre-exponential factors in, respectively, Eqs.6 and 9, for the i[th] regime of species $Y_j$. They are labelled by either $\Phi$ or $\Delta$. Also fitting parameters of these equations |
| $X$ | Reactant, $X = Y_0$ |
| $XY_v(q\Phi, uk)$ | Photothermal reaction involving a reactant and $v$ product, linked by $q$ photochemical ($\Phi$) and $u$ thermal ($k$) reaction-steps |
| $Y_j$ | Photoproducts, $j \neq 0$ $(1 \leq j \leq n_{sp})$ |
| $Y_{j'}$ | Photoproduct linked to $Y_j$ by a photochemical or a thermal reaction-step, with $j \neq j'$ |

*Molecules* **2024**, *29*, x FOR PEER REVIEW 24 of 26## References

1. Maafi, M. Excitation-wavelength dependent photochemistry. *Photochem*. **2024**, 4, 233-270. doi. 10.3390/photochem4020015
2. Mauser, H., Gauglitz, G., Compton, R.G., and Hancock, G. (Eds.). Comprehensive chemical kinetics, photokinetics: theoretical fundamentals and applications. Vol. 36. Elsevier, Amesterdam. **1998**. ISBN-13. 978-0444825360
3. Tonnesen, H.H. (Ed.). Photostability of Drugs and Drug Formulations. CRC Press, Boca Raton. **2004**. ISBN. 0-415-30323--0
4. Griesbeck, A., Oelgemoller, M.; and Ghetti, F. (ed.). Handbook of Organic Photochemistry and Photobiology (3rd Ed.). CRC Press. Boca Raton, New York, London. **2012**. ISBN 9781439899335
5. Lente, G. Deterministic Kinetics in Chemistry and Systems Biology: The Dynamics of Complex Reaction Networks. Springer, Cham, Heidelberg, New York, Dordrecht, London, **2015**. doi. 10.1007/978-3-319-15482-4
6. Yamada, H., and Yagai, S. (ed.). Light-Active Functional Organic Materials. Jenny Stanford Publishing. **2019**. ISBN. 9789814800150
7. Williams, J.D., and Kappe, C.O. Recent advances toward sustainable flow photochemistry. *Current Opinion in Green and Sustainable Chemistry*. **2020**, 25, 100351. doi. 10.1016/j.cogsc.2020.05.001
8. Balzani, V., Ceroni, P., Juris, A. Photochemistry and Photophysics: Concepts, Research, Applications (2nd Ed.). Wiley-VCH, **2024**. ISBN. 978-3-527-84408-1
9. McFadden, M.E, Barber, R.W., Overholts, A.C., Robb, M.J. Naphthopyran molecular switches and their emergent mechanochemical reactivity. *Chem. Sci*. **2023**, 14, 10041-10067. doi. 10.1039/D3SC03729K
10. Sun, Y., McFadden, M.E., Osler, S.K., Barber, R.W., Robb, M.J. Anomalous photochromism and mechanochromism of a linear naphthopyran enabled by a polarizing dialkylamine substituent. *Chem. Sci*. **2023**, 14, 10494-10499. doi. 10.1039/D3SC03790H
11. Minkin, V.I. Photo-, thermo-, solvato-, and electrochromic spiroheterocyclic compounds. *Chem. Rev*. **2004**, 104, 2751–2776. doi. 10.1021/cr020088u
12. Li, C., Zhang, Y., Hu, J., Cheng, J.J., Liu, S.Y. Reversible three-state switching of multicolor fluorescence emission by multiple stimuli modulated FRET processes within thermoresponsive polymeric micelles. *Angew. Chem. Int. Ed*. **2010**, 122, 5246–5250. doi. 10.1002/ange.201002203
13. Samai, S., Bradley, D.J., Choi, T.L.Y., Yan, Y. Temperature-Dependent Photoisomerization Quantum Yields for Azobenzene-Modified DNA. *J. Phys. Chem. C*. **2017**, 121, 6997-7004. doi. 10.1021/acs. jpcc.6b12241
14. Li, C., and Liu, S. Polymeric assemblies and nanoparticles with stimuli-responsive fluorescence emission characteristics. *Chem. Commun*. **2012**, 48, 3262–3278. doi. 10.1039/C2CC17695E
15. Xiong, Y., Jentzsch, A.V., Osterrieth, J.W.M., Sezgin, E., Sazanovich, I.V., Reglinski, K., Galiani, S., Parker, A.W., Eggeling, C., Anderson, H.L. Spironaphthoxazine switchable dyes for biological imaging. *Chem. Sci*. **2018**, 9, 3029-3040. doi. 10.1039/C8SC00130H
16. Fagan, A., Bartkowski, M, Giordani, S. Spiropyran-Based Drug Delivery Systems. *Front. Chem*. **2021**, 9, 720087. doi. 10.3389/fchem.2021.720087
17. Aillet, T., Loubiere, K., Dechy-Cabaret, O., and Prat, L. Accurate measurement of the photon flux received inside two continuous flow microphotoreactors by actinometry. *Int. J. Chem. React. Eng*. **2014**, 12, 1–13. doi. 10.1515/ijcre-2013-0121
18. Su, C.D., Shi, Y.Y., and Gao, J. Synthesis and properties of photochromic polymer contain spiro-oxazine induced by ultraviolet light. *Soft Materials*. **2023**, 21, 1-13. doi. 10.1080/1539445X.2022.2108842
19. Yan, J., Zhao, L.X., Li, C., Hu, Z., Zhang, G., Chen, Z., Chen, T., Huang, Z., Zhu, J., Zhu, M. Optical nanoimaging for block copolymer self-assembly. *J. Am. Chem. Soc*. **2015**, 137, 2436–2439. doi. 10.1021/ja512189a
20. Bukreeva, T., Barachevsky, V., Venidiktova, O., Krikunova, P., Pallaeva, T. Recent development of photochromic polymer capsules for smart materials. *Materials Today Communications*. **2024**, 38, 107769, doi. 10.1016/j.mtcomm.2023.107769
21. Zitzmann, M., Fröhling, M., Dube, H. Gain of Function Recyclable Photoswitches: Reversible Simultaneous Substitution and Photochromism Generation. *Angew. Chem. Int. Ed. Engl*. **2024**, 63, e202318767. doi. 10.1002/anie.202318767
22. Willis, G.L., Boda, J., Freelance, C.B. Polychromatic Light Exposure as a Therapeutic in the Treatment and Management of Parkinson's Disease: A Controlled Exploratory Trial. *Front. Neurol*. **2018**, 9, 741. doi: 10.3389/fneur.2018.00741
23. Tarr, M., Zito, P. Photochemistry of nanomaterials: environmental impacts. American Chemical Society Publications, **2022**. doi. 10.1021/acsinfocus.7e5012
24. Ge, L., Wang, S., Halsall, C., Li, X., Bai, D., Cao, S., Zhang, P. New insights into the environmental photochemistry of common-use antibiotics in ice and in water: A comparison of kinetics and influencing factors. *Emerging Contaminants*. **2024**, 10, 100382. doi. 10.1016/j.emcon.2024.100382
25. Zhu, M.Q.; Zhu, L.; Han, J.J.; Wu, W.; Hurst, J.K.; Li, A.D. Spiropyran-based photochromic polymer nanoparticles with optically switchable luminescence. *J. Am. Chem. Soc*. **2006**, 128, 4303–4309. doi. 10.1021/ja0567642
26. Chen, Q., Feng, Y., Zhang, D., Zhang, G., Fan, Q., Sun, S., Zhu, D. Light-triggered self-assembly of a spiropyran-functionalized dendron into nano-/micrometer-sized particles and photoresponsive organogel with switchable fluorescence. *Adv. Funct. Mater*. **2010**, 20, 36–42. doi. 10.1002/adfm.200901358
27. Grzelczak, M.; Vermant, J.; Furst, E.M.; Liz-Marzán, L.M. Directed self-assembly of nanoparticles. *ACS Nano*. **2010**, 4, 3591–3605. doi. 10.1021/nn100869j
28. Hassaan, M.A., El-Nemr, M.A., Elkatory, M.R., Ragab, S., Niculescu, V.C., El Nemr, A. Principles of Photocatalysts and Their Different Applications: A Review. *Top. Curr. Chem. (Z)*. **2023**, 381, 31. doi. 10.1007/s41061-023-00444-7
29. Udayabhaskararao, T., Kundu, P.K., Ahrens, J., Klajn, R. Reversible Photoisomerization of Spiropyran on the Surfaces of Au25 Nanoclusters. *ChemPhysChem*. **2016**, 17, 1805-1809. doi. 10.1002/cphc.201500897